\documentclass[11pt,a4paper]{article}
\usepackage{epsfig,amsmath,amssymb,scalefnt,cite,mcite}

\long\def\symbolfootnote[#1]#2{\begingroup%
\def\thefootnote{\fnsymbol{footnote}}\footnote[#1]{#2}\endgroup}

\let\OLDthebibliography\thebibliography
\renewcommand\thebibliography[1]{
  \OLDthebibliography{#1}
  \setlength{\parskip}{0pt}
  \setlength{\itemsep}{0pt plus 0.3ex}
}

\newcommand{\abbrev}{\rm\scalefont{.9}}
\newcommand{\zwat}{{\abbrev ZWA}}
\newcommand{\sm}{{\abbrev SM}}

\newcommand{\lhc}{{\abbrev LHC}}

\newcommand{\cms}{{\abbrev cms}}
\newcommand{\lhchxswg}{{\abbrev LHC-HXSWG}}

\newcommand{\smallz}{{\scriptscriptstyle Z}} 
\newcommand{\smallw}{{\scriptscriptstyle W}} %
\newcommand{\smallv}{{\scriptscriptstyle V}} %
\newcommand{\smallH}{{\scriptscriptstyle H}} %

\newcommand{\smallga}{{\scriptscriptstyle \gamma}} %
\newcommand{\mz}{m_\smallz}
\newcommand{\mv}{m_\smallv}
\newcommand{\mzz}{m_{\smallz\smallz}}
\newcommand{\mvv}{m_{\smallv\smallv}}
\newcommand{\mww}{m_{\smallw\smallw}}
\newcommand{\mgaga}{m_{\smallga\smallga}}
\newcommand{\mw}{m_\smallw}
\newcommand{\mh}{m_\smallH}

\newcommand{\mj}{m_{4j}}

\newcommand{\zh}{{\smallz\smallH}}

\newcommand{\zzz}{{\smallz\smallz\smallz}}
\newcommand{\zvv}{{\smallz\smallv\smallv}}
\newcommand{\zww}{{\smallz\smallw\smallw}}

\newcommand{\nunuzz}{{\nu\bar\nu\smallz\smallz}}
\newcommand{\nunuww}{{\nu\bar\nu\smallw\smallw}}
\newcommand{\nunuvv}{{\nu\bar\nu\smallv\smallv}}
\newcommand{\zgaga}{{\smallz\smallga\smallga}}

\newcommand{\GaH}{\Gamma_\smallH}

\newcommand{\zwa}{{\rm\scriptscriptstyle ZWA}}

\newcommand{\eqn}[1]{Eq.\,(\ref{#1})}

\newcommand{\fig}[1]{Fig.\,\ref{#1}}
\newcommand{\figs}[1]{Figs.\,\ref{#1}}
\newcommand{\tab}[1]{Tab.\,\ref{#1}}
\newcommand{\sct}[1]{Section~\ref{#1}}

\newcommand{\citere}[1]{Ref.~\cite{#1}}
\newcommand{\citeres}[1]{Refs.~\cite{#1}}

\begin{document}

\begin{titlepage}

{\flushright{
        \begin{minipage}{2.5cm}
          DESY 15-043
        \end{minipage}        }

}
\renewcommand{\thefootnote}{\fnsymbol{footnote}}
\vskip 2cm
\begin{center}
{\LARGE\bf Off-shell effects in Higgs decays to
heavy gauge bosons and signal-background interference
in Higgs decays to photons at a linear collider}
\vskip 1.0cm
{\Large  Stefan Liebler}
\vspace*{8mm} \\
{\sl 
DESY, Notkestra\ss e 85, \\
22607 Hamburg, Germany}
\end{center}
\symbolfootnote[0]{{\tt Talk presented at the International Workshop on Future Linear Colliders (LCWS14), Belgrade, Serbia, 6-10 October 2014.}}

\vskip 0.7cm

\begin{abstract}
We discuss off-shell contributions in Higgs decays to heavy
gauge bosons $H\rightarrow VV^{(*)}$ with $V\in\lbrace Z,W\rbrace$ for a standard model
(\sm{}) Higgs boson for both dominant
production processes $e^+e^-\rightarrow ZH\rightarrow ZVV^{(*)}$ and
$e^+e^-\rightarrow \nu\bar\nu H\rightarrow \nu\bar\nu VV^{(*)}$ 
at a (linear) $e^+e^-$ collider. Dependent on the centre-of-mass energy
off-shell effects are sizable and important for the understanding
of the electroweak symmetry breaking mechanism.
Moreover we shortly investigate the effects of the signal-background
interference in $H\rightarrow \gamma\gamma$ decays for
the Higgsstrahlung initiated process $e^+e^-\rightarrow Z\gamma\gamma$,
where we report a similar shift in the invariant mass peak of the 
two photons as found for the \lhc{}. For both effects we
discuss the sensitivity to the total Higgs width.
\end{abstract}
\vfill
\end{titlepage}    

\setcounter{footnote}{0}


\section{Introduction}
\label{sec:intro}

After the discovery of a scalar boson at the Large Hadron Collider (\lhc{})
\cite{Aad:2012tfa,Chatrchyan:2012ufa}
a major task of particle physics is the precise determination of its couplings
and its mass in order to reveal the mechanism of electroweak symmetry breaking
and to determine whether it is the standard model (\sm{}) Higgs boson.
Recently two interesting effects to test the underlying nature 
were discussed at the \lhc{}, a measurement of off-shell Higgs boson decays into heavy gauge
bosons as well as a shift in the invariant mass peak of the two photons
in Higgs boson decays to photons, the latter due to signal-background interference terms. Both
-- under certain theoretical assumptions -- are sensitive to the total Higgs width.
It is timely to discuss the two effects for a (linear) $e^+e^-$ collider, not only
for what concerns constraints on the Higgs width, but also to elaborate on
the phenomenological consequences of the effects. We subsequently first discuss
off-shell Higgs boson decays to heavy gauge bosons and afterwards
the signal-background interference for Higgs boson decays to photons.

\section{Off-shell effects in $H\rightarrow VV^{(*)}$}
\label{sec:offshell}

In this section we want to discuss off-shell effects in
Higgs boson decays $H\rightarrow VV^{(*)}$ with $V\in\lbrace Z,W\rbrace$
at an $e^+e^-$ collider, where we
focus on both production processes $e^+e^-\rightarrow ZH$
and $e^+e^-\rightarrow \nu\bar\nu H$. For a more detailed
discussion we refer to \citere{Liebler:2015aka}, where also
the relation between the mass and the total
width of the Higgs boson and the complex pole of the
propagator is presented.
In accordance to the \lhc{} Higgs cross section working group
(\lhchxswg{})~\cite{Dittmaier:2011ti}
we choose the total width $\GaH^{\sm}=4.07\cdot 10^{-3}$\,GeV
for a \sm{} Higgs boson with mass $\mh=125$\,GeV in the subsequent discussion.

\begin{figure}[ht]
\begin{center}
\begin{tabular}{cc}
\includegraphics[width=0.3\textwidth]{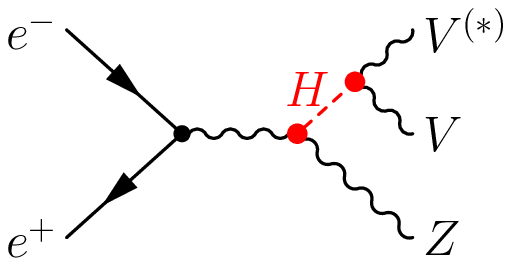} &
\includegraphics[width=0.3\textwidth]{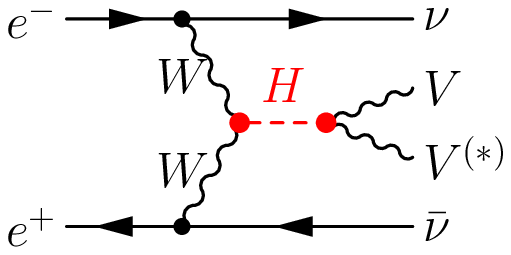} \\[-0.5cm]
 (a) & (b)
\end{tabular}
\end{center}
\vspace{-0.6cm}
\caption{Feynman diagrams for the two dominant production processes
(a) $e^{+}e^{-} \rightarrow ZH$ followed by $H\rightarrow VV^{(*)}$ and
(b) $e^{+}e^{-} \rightarrow \nu\bar\nu H$ followed by $H\rightarrow VV^{(*)}$.}
\label{fig:feynmanZZ} 
\end{figure}

Following \citeres{Kauer:2012hd}
for the  case of the \lhc{}, it is well established that the kinematic
region, where the invariant mass of the two gauge bosons $\mvv$
in $H\rightarrow VV^{(*)}$ exceeds twice the gauge boson masses $2\mv$, contributes a sizable fraction
of several percent to the total inclusive cross section $gg\rightarrow H\rightarrow VV^{(*)}$.
A combination of on- and off-shell effects was identified to allow
to constrain the total Higgs width~\cite{Caola:2013yja},
however under strong theoretical assumptions~\cite{Englert:2014aca}.
Obtained experimental bounds from the CMS and the ATLAS collaboration
can be found in \citeres{CMS:2014ala},
where apart from the gluon fusion production process a similar effect for
vector-boson fusion was used.
Recently a discussion of the off-shell effects for an $e^+e^-$ collider
was carried out in \citere{Liebler:2015aka} with many analogies to
the \lhc{} discussion. Subsequently we shortly summarize
the findings:
The large off-shell contribution to the total process
$e^+e^-\rightarrow ZH$ and $e^+e^-\rightarrow \nu\bar\nu H$
with the Higgs boson decays $H\rightarrow WW^{(*)}$ and $H\rightarrow ZZ^{(*)}$
(see \fig{fig:feynmanZZ} for Feynman diagrams)
can be understood as inadequacy of the zero-width approximation (\zwat{})
in the description of the process. Using it for the production and
decay part of the process one can define~\cite{Kauer:2012hd}
\begin{align}
\label{eq:ZWA}
\left(\frac{d\sigma^{\zvv}_{\zwa}}{d\mvv }\right)
=\sigma^{\zh}(\mh)\frac{2\mvv}{(\mvv^2-\mh^2)^2+(\mh\GaH)^2}\frac{\mh\Gamma_{H\rightarrow
VV}(\mh)}{\pi}\quad,
\end{align}
which we later compare to the description
given by the off-shell production cross section according to \cite{Kauer:2012hd,Liebler:2015aka}
\begin{align}
\left(\frac{d\sigma^\zvv_{\text{off}}}{d\mvv}\right)
=\sigma^{\zh}(\mvv)\frac{2\mvv}{(\mvv^2-\mh^2)^2 + 
(\mh\GaH)^2}\frac{\mvv\Gamma_{H \rightarrow VV}(\mvv)}{\pi}\quad.
\label{eq:sigmaoff}
\end{align}

For two processes additional comments are on order:
For $e^+e^-\rightarrow \nu\bar\nu W^+W^-$ apart from the contribution
which involves an $s$-channel Higgs boson in the off-shell region
corresponding to the approximation
given by \eqn{eq:sigmaoff}, also the $t$-channel Higgs boson exchange
between the gauge bosons is of relevance and thus added to the
Higgs boson induced contributions.
Secondly for $e^+e^-\rightarrow ZH\rightarrow ZZZ$
it is a priori unclear which two out of three $Z$ bosons
originate from an intermediate Higgs boson.
We thus also average over the three possible invariant
mass combinations $\mvv$, which however induces on-shell
Higgs boson events to contribute in the off-shell region $\mzz>2\mz$.
For what concerns the gauge invariance of our discussion,
the inclusion of higher order contributions or initial-state radiation
we refer to \citere{Liebler:2015aka}.

\begin{figure}[htp]
\begin{center}
\begin{tabular}{cc}
\includegraphics[width=0.4\textwidth]{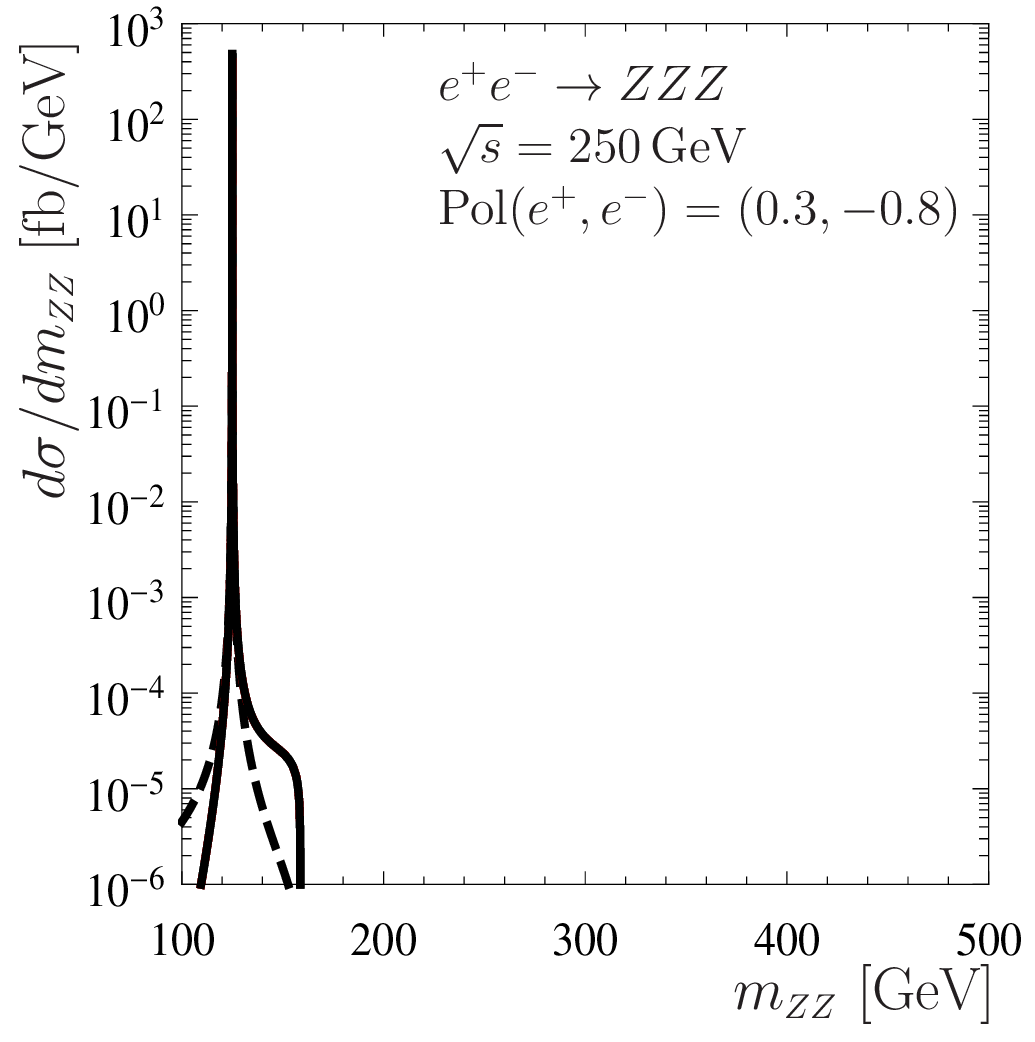} &
\includegraphics[width=0.4\textwidth]{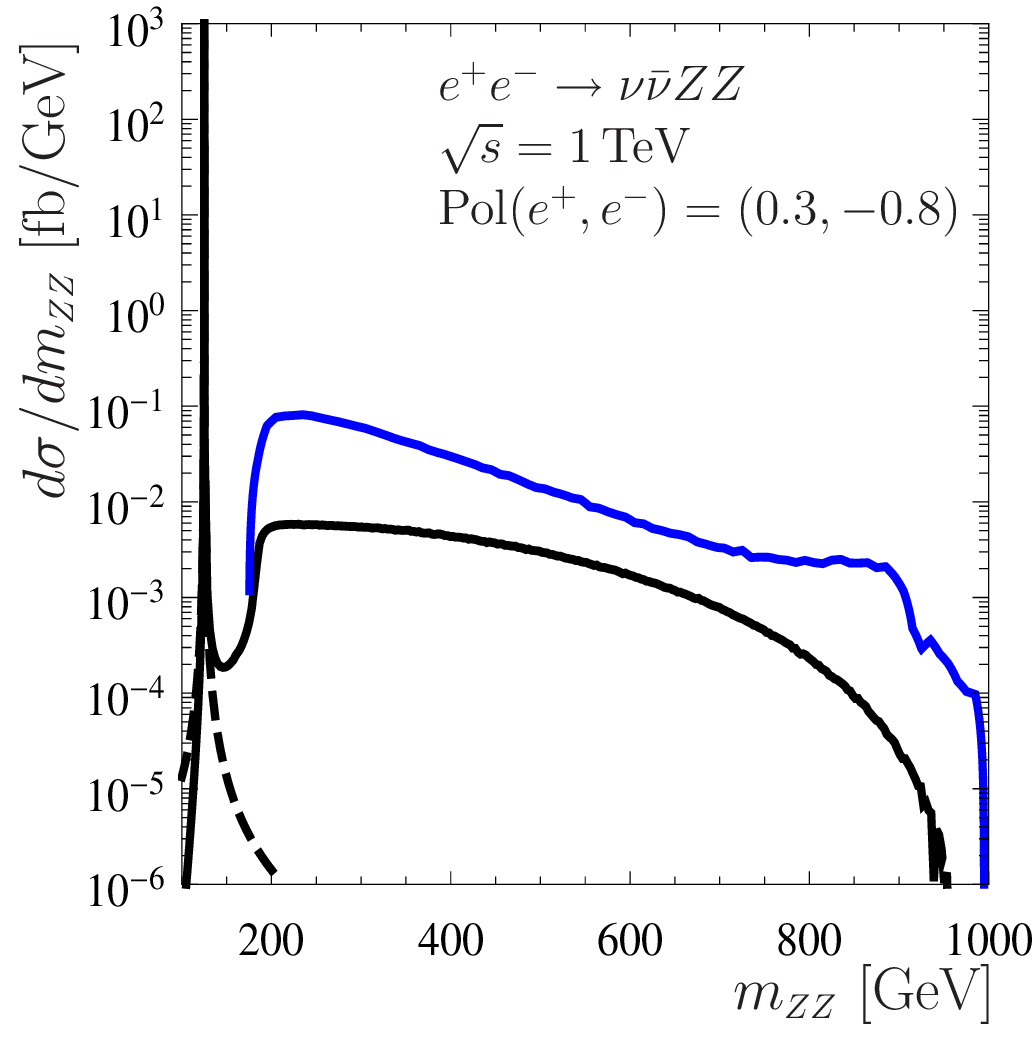}\\[-0.5cm]
 (a) & (b) \\
\includegraphics[width=0.4\textwidth]{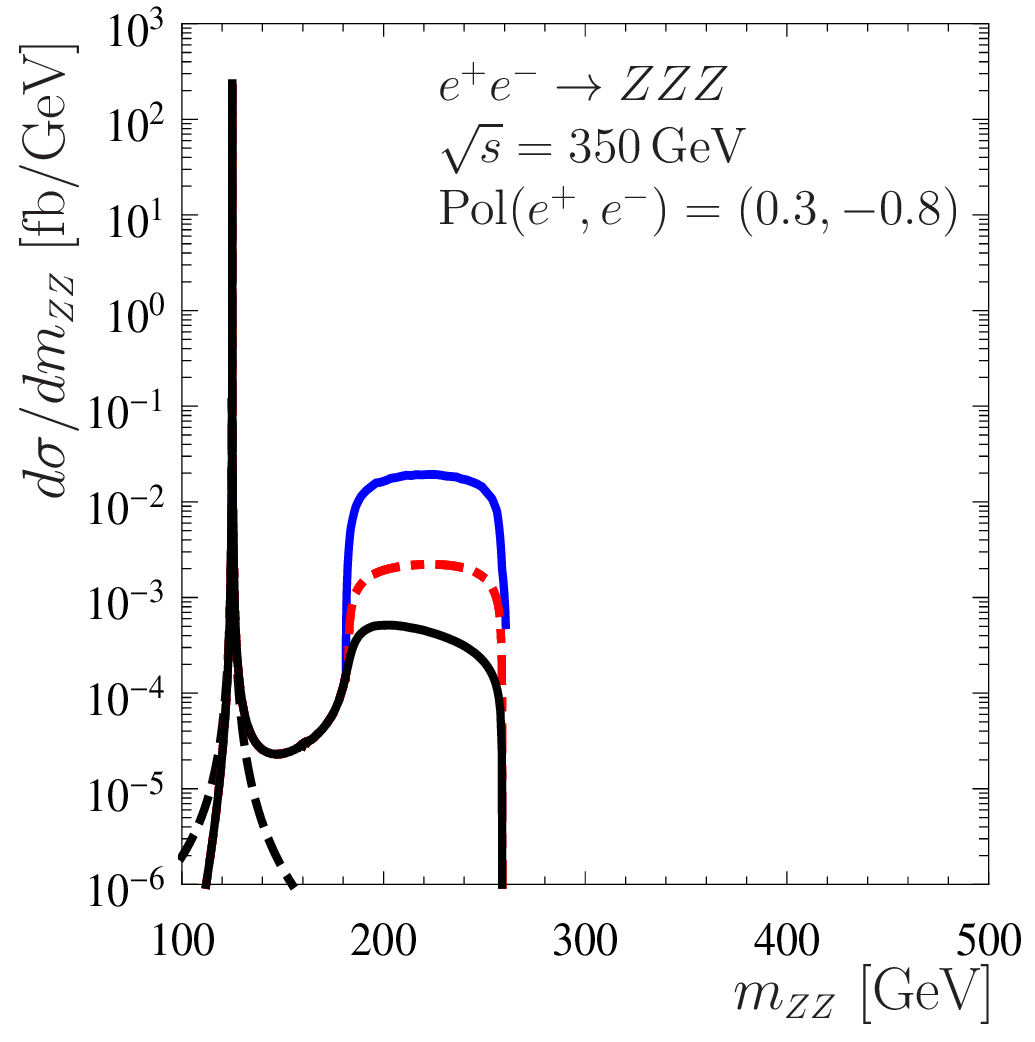} &
\includegraphics[width=0.4\textwidth]{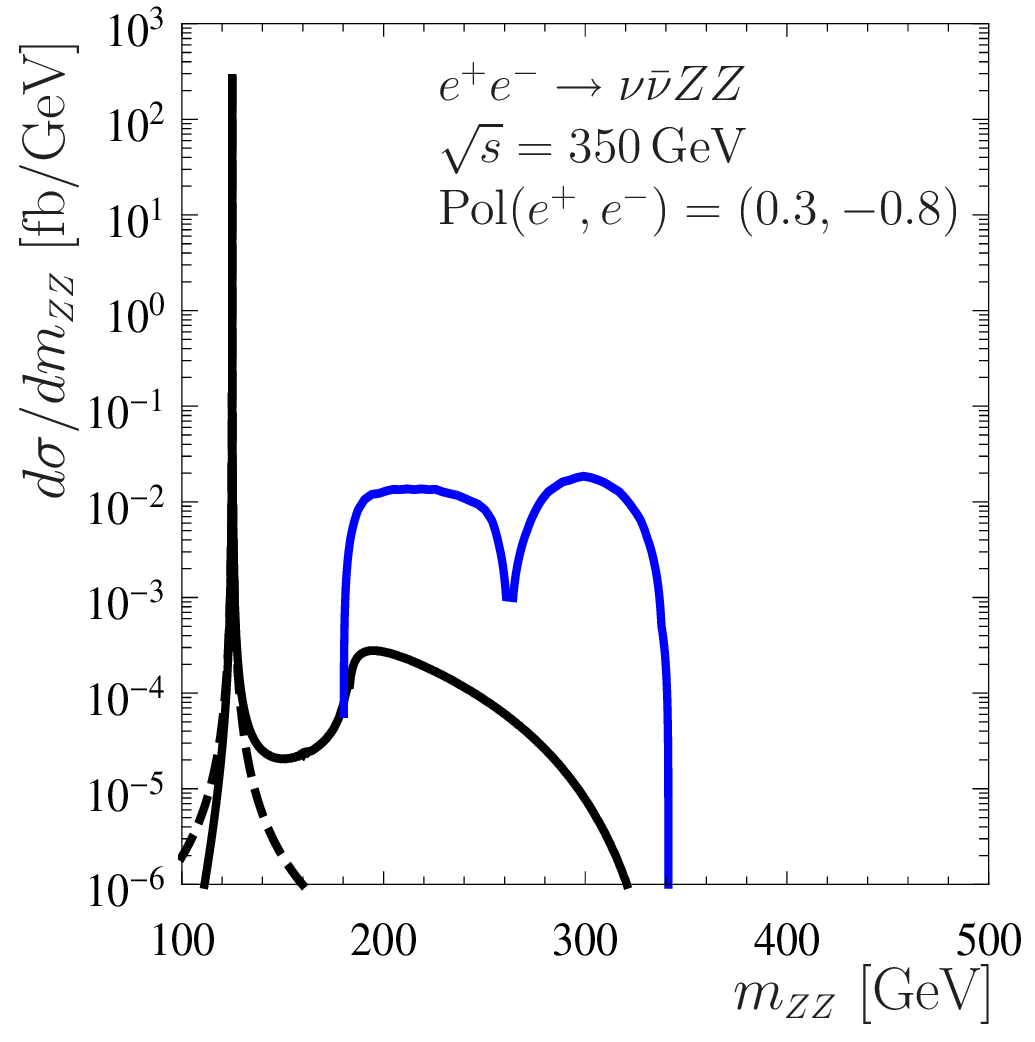}\\[-0.5cm]
 (c) & (d) \\
\includegraphics[width=0.4\textwidth]{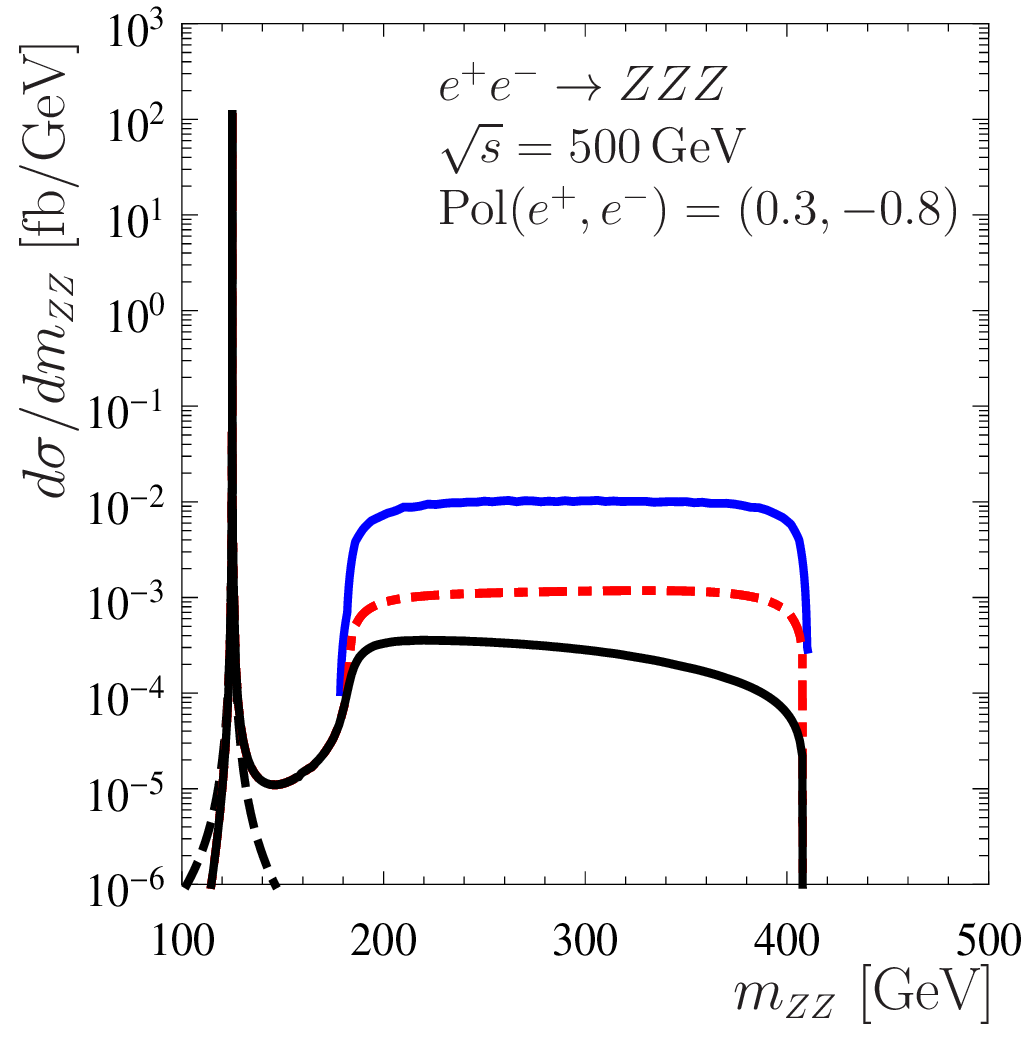} &
\includegraphics[width=0.4\textwidth]{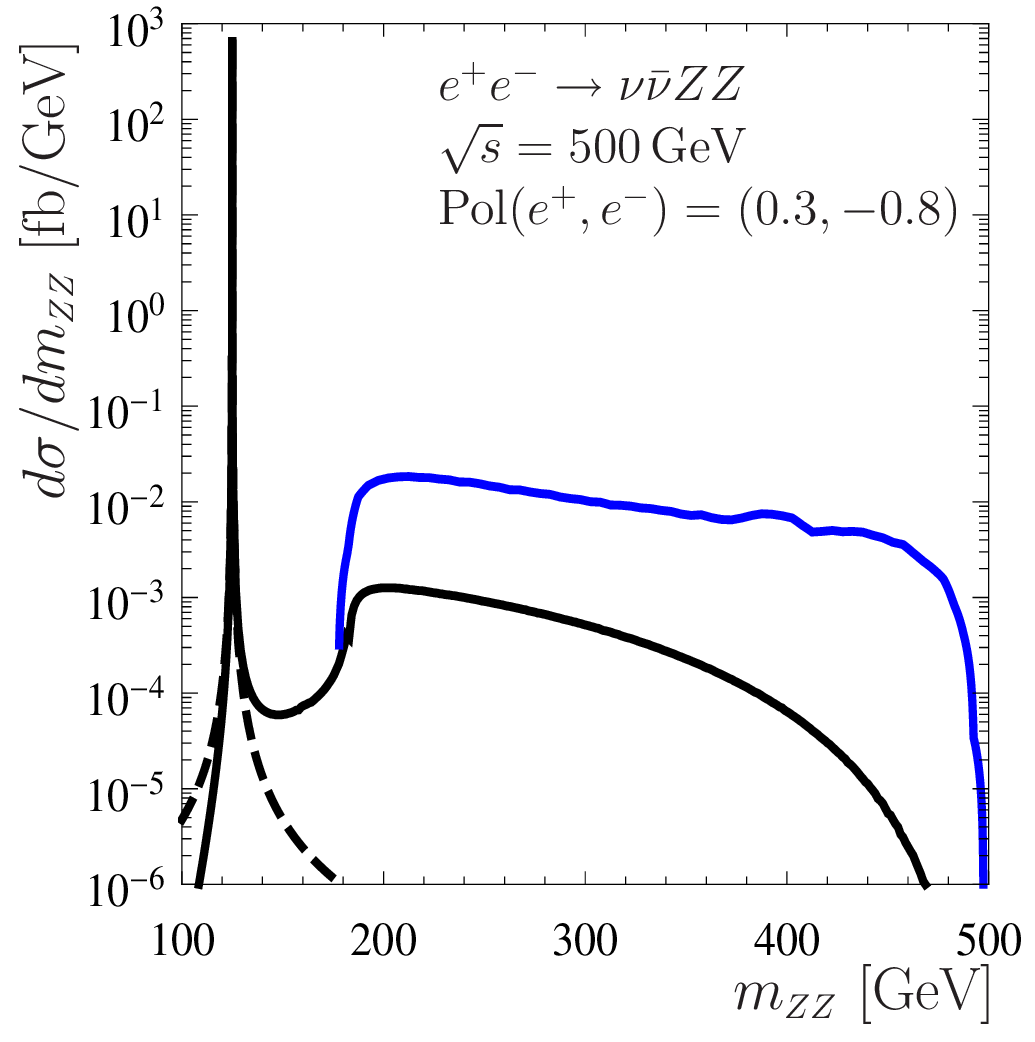}\\[-0.5cm]
 (e) & (f) \\
\end{tabular}
\end{center}
\vspace{-0.6cm}
\caption{$d\sigma/d\mzz$ in fb/GeV as a function of $\mzz$ in GeV
according to \eqn{eq:ZWA} (\zwat{}) (black, dashed)
and \eqn{eq:sigmaoff} (black, solid) for
(a,c,e) $e^+e^-\rightarrow ZH\rightarrow ZZZ$ for
\cms{} energies $\sqrt{s}=250,\, 350,\, 500$\,GeV (top to bottom)
and (b,d,f) $e^+e^-\rightarrow\nu\bar\nu H\rightarrow \nu\bar\nu ZZ$
for \cms{} energies $\sqrt{s}=1000,\, 350,\, 500$\,GeV (top to bottom)
with a fixed polarisation Pol$(e^+,e^-)=(0.3,-0.8)$.
The calculation of 
$e^+e^-\rightarrow ZH\rightarrow ZZZ$ with an average over the $ZZ$ pairs
is presented by the red, dot-dashed curve.
The complete calculation $e^+e^-\rightarrow ZZZ/\nu\bar\nu ZZ$
is depicted by the blue curve.
}
\label{fig:ZZprocess} 
\end{figure}

\begin{figure}[htp]
\begin{center}
\begin{tabular}{cc}
\includegraphics[width=0.4\textwidth]{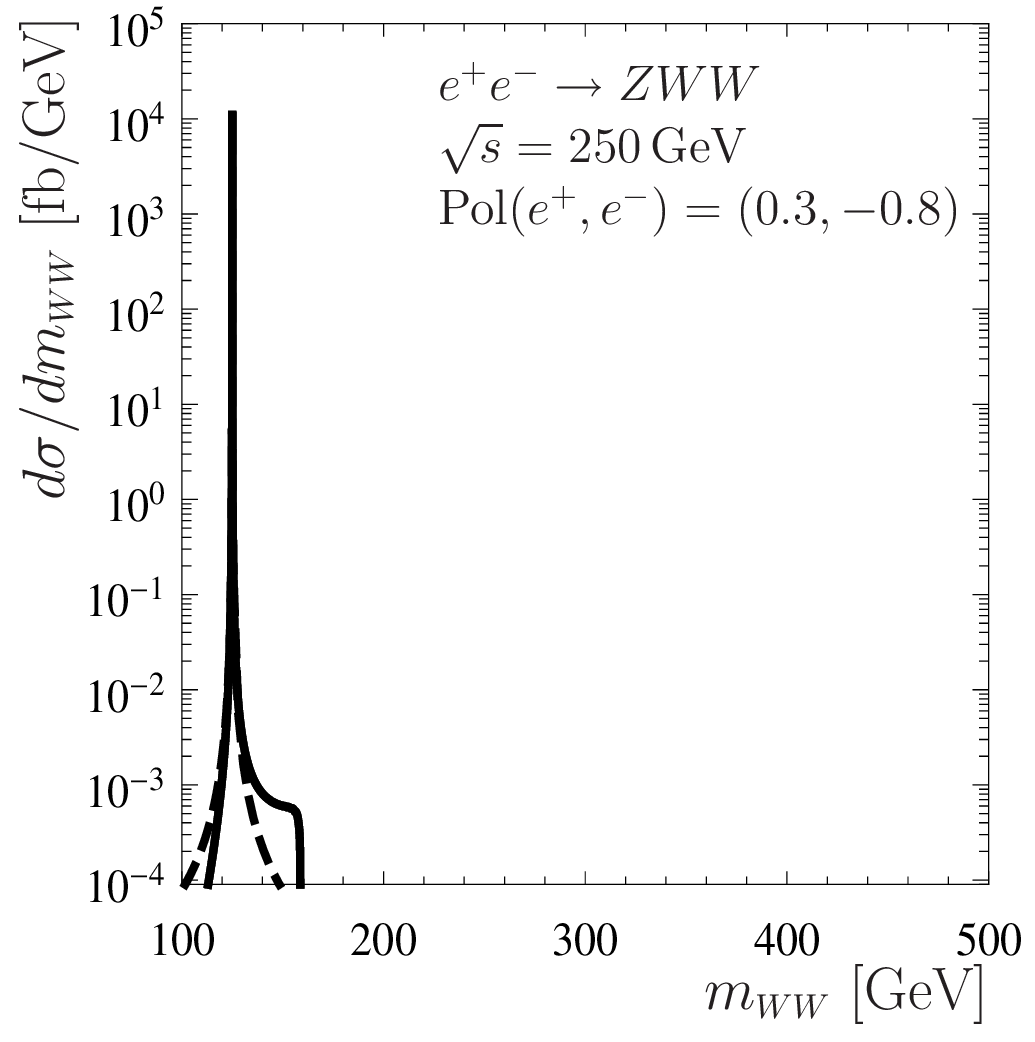} &
\includegraphics[width=0.4\textwidth]{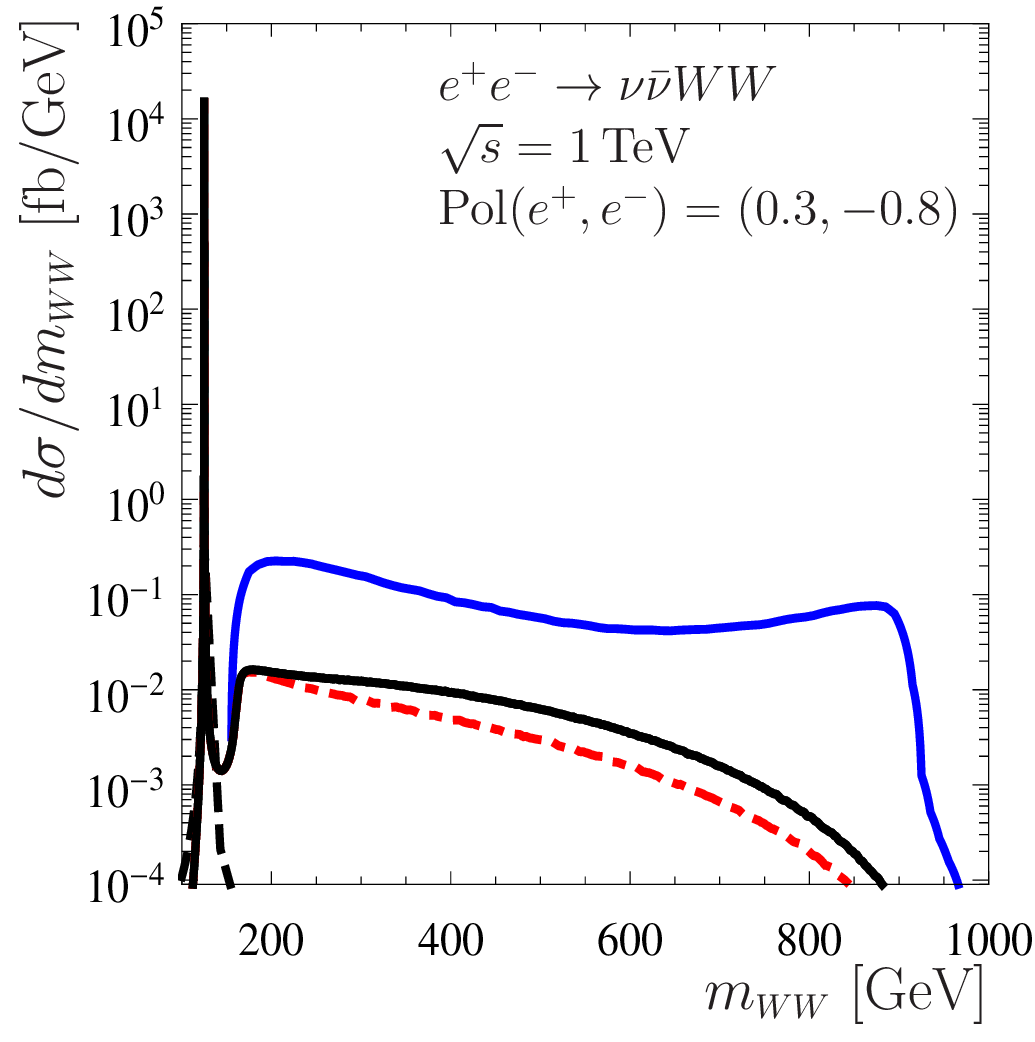}\\[-0.5cm]
 (a) & (b) \\
\includegraphics[width=0.4\textwidth]{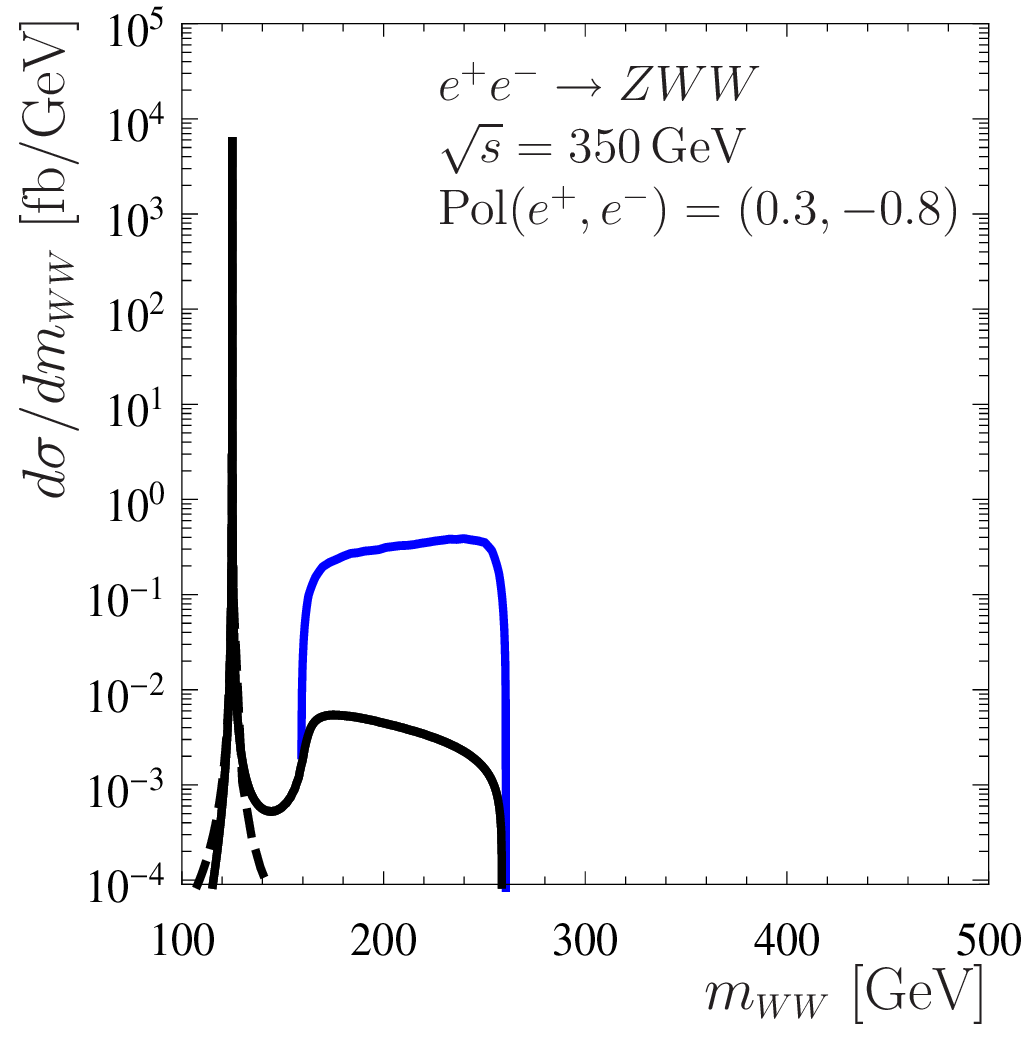} &
\includegraphics[width=0.4\textwidth]{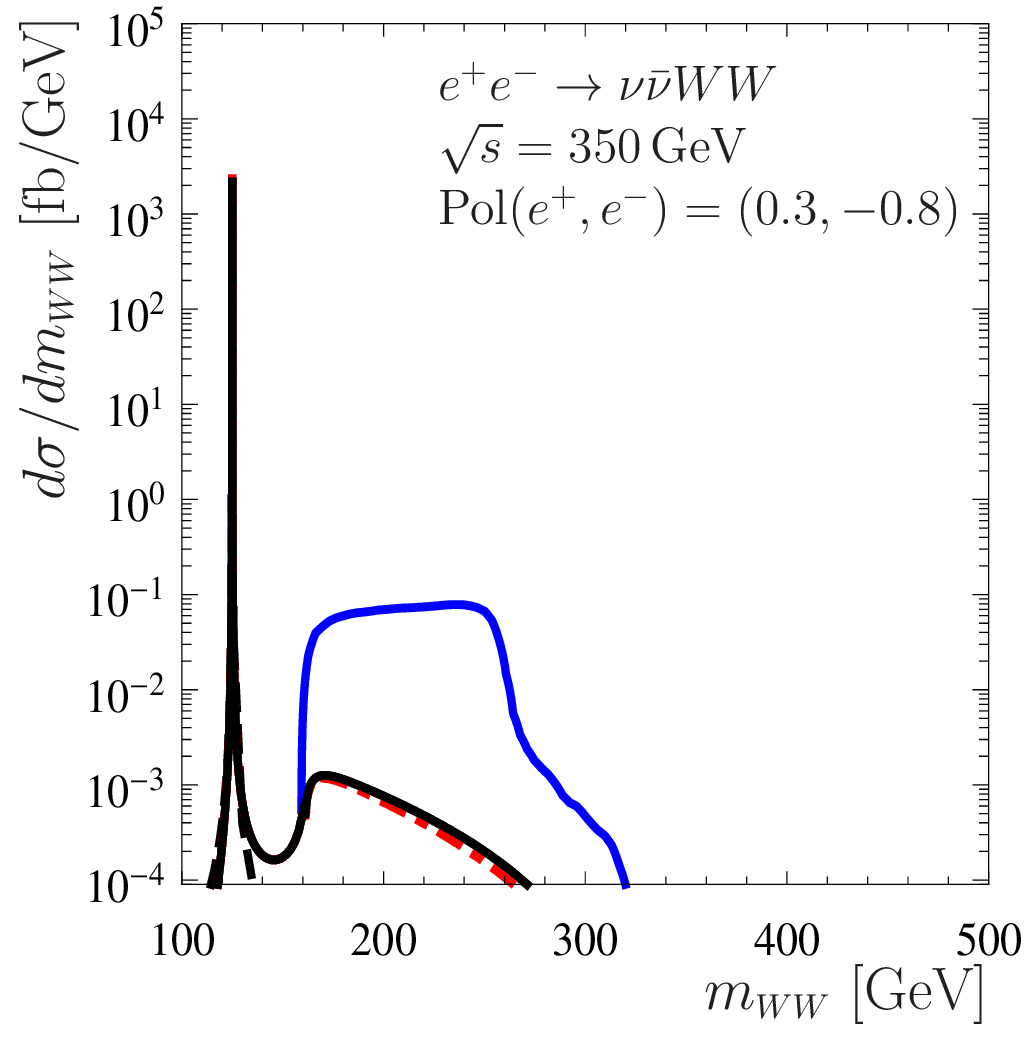}\\[-0.5cm]
 (c) & (d) \\
\includegraphics[width=0.4\textwidth]{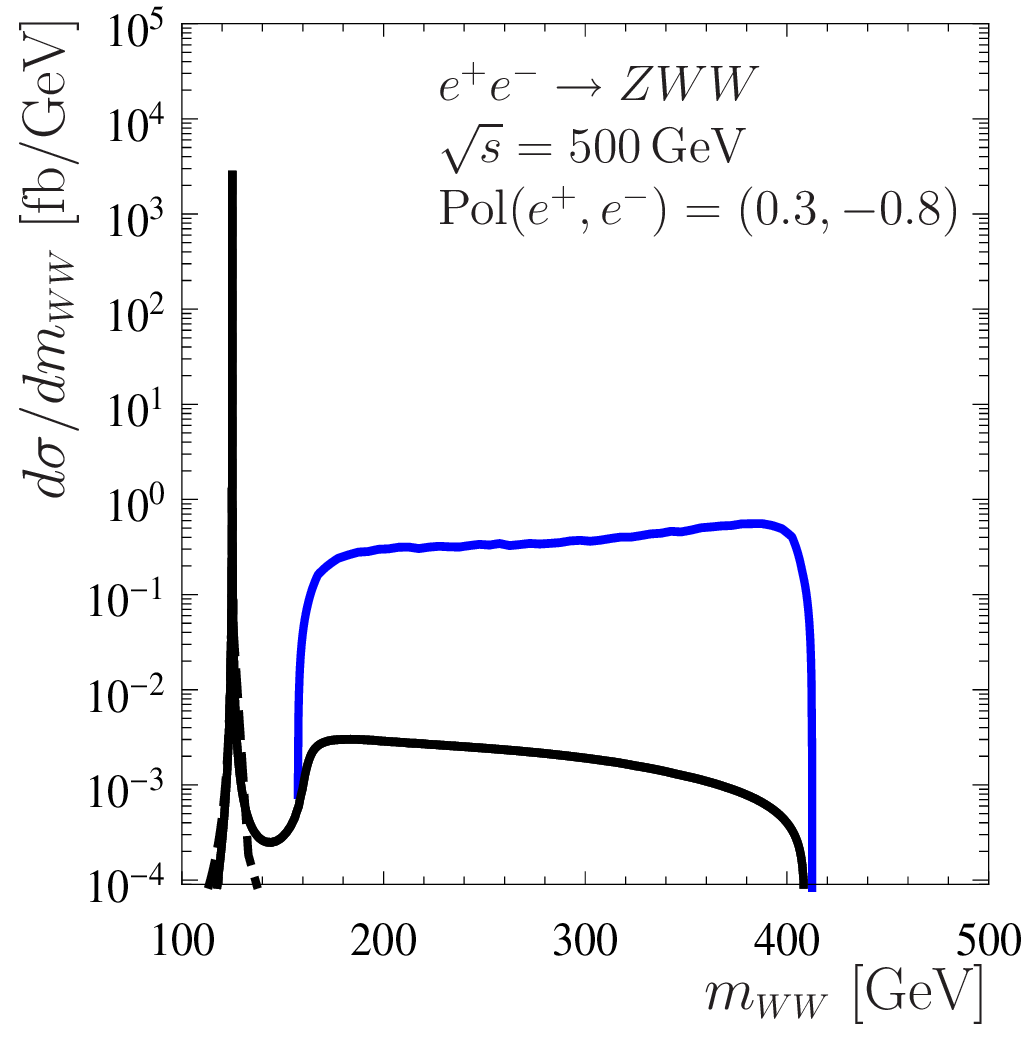} &
\includegraphics[width=0.4\textwidth]{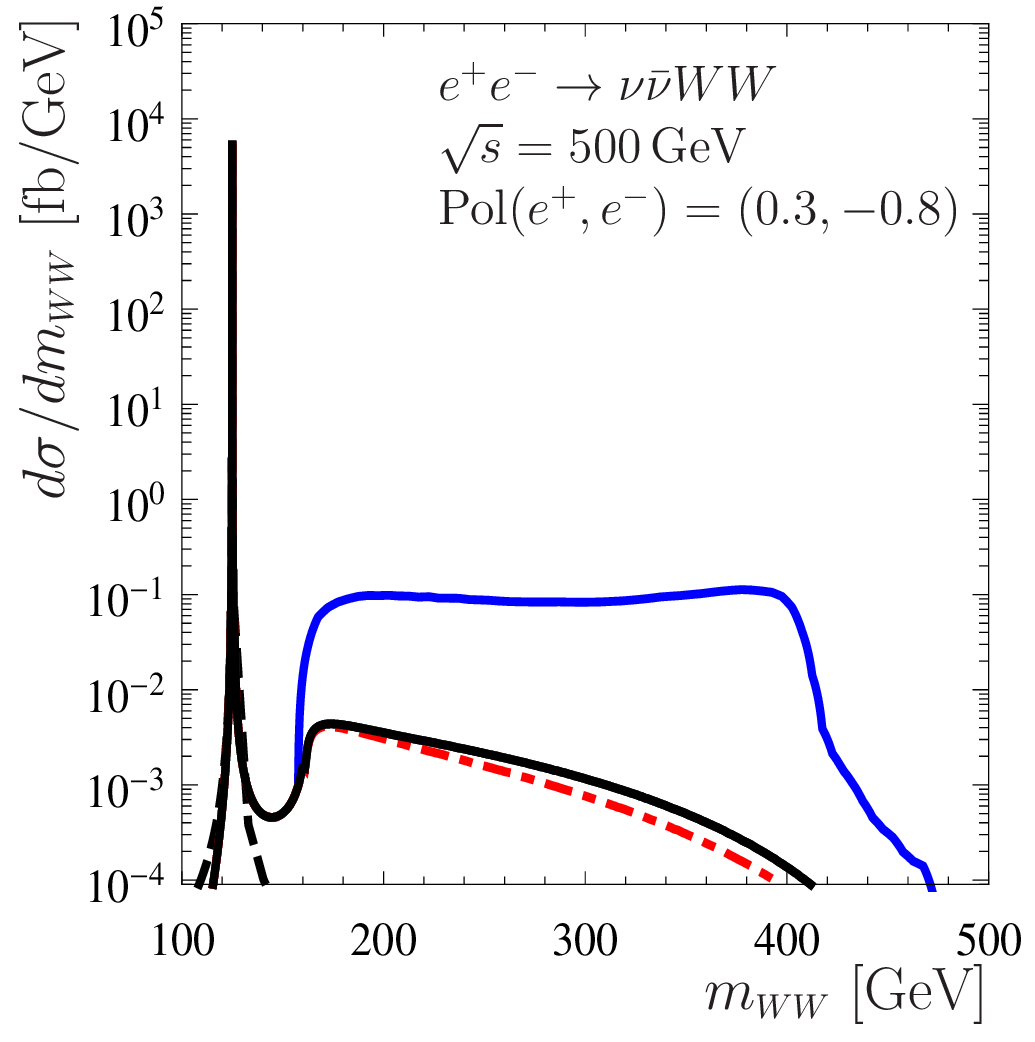}\\[-0.5cm]
 (e) & (f) \\
\end{tabular}
\end{center}
\vspace{-0.6cm}
\caption{$d\sigma/d\mww$ in fb/GeV as a function of $\mww$
in GeV according to \eqn{eq:ZWA} (\zwat{}) (black, dashed)
and \eqn{eq:sigmaoff} (black, solid) for
(a,c,e) $e^+e^-\rightarrow ZH\rightarrow ZWW$ for \cms{}
energies $\sqrt{s}=250,\, 350,\, 500$\,GeV (top to bottom)
and (b,d,f) $e^+e^-\rightarrow \nu\bar\nu H\rightarrow \nu\bar\nu WW$
for \cms{} energies $\sqrt{s}=1000,\, 350,\, 500$\,GeV (top to bottom)
with a fixed polarisation Pol$(e^+,e^-)=(0.3,-0.8)$.
We add both $s$- and $t$-channel Higgs induced contributions for
$e^+e^-\rightarrow \nu\bar\nu WW$ as red, dot-dashed curve.
The complete differential cross section including background diagrams
yielding the same final state is depicted as blue curve.
}
\label{fig:WWprocess} 
\end{figure}

We show the differential cross sections $d\sigma / d\mzz$
for $e^+e^-\rightarrow ZH$ and $e^+e^-\rightarrow \nu\bar\nu H$
followed by $H\rightarrow ZZ^{(*)}$ 
for different centre-of-mass (\cms{}) energies $\sqrt{s}=250,\, 350,\, 500$\,GeV and $1$\,TeV,
but a fixed polarisation of the initial state being Pol$(e^+,e^-)=(0.3,-0.8)$ in \fig{fig:ZZprocess}.
Our results are obtained using {\tt FeynArts}, {\tt FormCalc}~\cite{Hahn:1998yk}
and {\tt MadGraph5\_aMC@NLO}~\cite{Alwall:2011uj}.
As indicated for $e^+e^-\rightarrow ZZZ$ we average over the three possible
invariant mass combinations of $ZZ$ pairs presented by the red, dot-dashed
curve, but also present the result obtained by the usage of \eqn{eq:sigmaoff}.
Similarly, we show the differential cross sections for both production processes
followed by $H\rightarrow WW^{(*)}$ in \fig{fig:WWprocess}, where we
add the red, dot-dashed curve including the $t$-channel Higgs boson induced
contributions.
In both \figs{fig:ZZprocess} and \ref{fig:WWprocess} we also show 
the total differential cross section for the specific final state as blue curve,
thus including the contributions from background diagrams with the same final state.
For a detailed discussion of the background we refer to \citere{Liebler:2015aka}.

In order to quantify the relative importance of the off-shell signal contributions
we define
\begin{align}
\label{eq:deltaoff}
\Delta_{\text{off}}^\zvv=\frac{\sigma^\zvv_{\text{off}}(130\text{GeV},
\sqrt{s} - \mz)}{\sigma^\zvv_{\text{off}}}\qquad
\text{and}
\qquad
\Delta_{\text{off}}^\nunuvv=\frac{\sigma^\nunuvv_{\text{off}}(130\text{GeV},
\sqrt{s} )}{\sigma^\nunuvv_{\text{off}}}
\end{align}
with the inclusive cross section for a lower and upper bound of
masses $\mvv$
\begin{align}
\sigma_{X}(\mvv^d,\mvv^u)=\int_{\mvv^d}^{\mvv^u} d\mvv\left(\frac{d\sigma_X}{d\mvv}\right)\quad.
\label{eq:uplowsigma}
\end{align}
and $\sigma^\zvv_{\text{off}}=\sigma^\zvv_{\text{off}}(0,\sqrt{s}-\mz)$
as well as $\sigma^\nunuvv_{\text{off}}=\sigma^\nunuvv_{\text{off}}(0,\sqrt{s})$.
$\Delta_{\text{off}}$ is hardly sensitive to the boundary value between
on- and off-shell contributions at $130$\,GeV. We present the polarisation
independent values $\Delta_{\text{off}}$
as a function of the \cms{} energy in \tab{tab:ZZtable}.
\begin{table}[htb]
\begin{center}
\begin{tabular}{| c || c | c || c | c |}
\hline
$\sqrt{s}$ & $\sigma^\zzz_{\text{off}}$ & $\Delta_{\text{off}}^\zzz$ 
& $\sigma^\nunuzz_{\text{off}}$ & $\Delta_{\text{off}}^\nunuzz$
\\\hline\hline
$250$\,GeV & $(3.12)3.12$\,fb   & $(0.03)0.03$\,\% & $0.490$\,fb & $0.12$\,\%\\\hline
$300$\,GeV & $(2.40)2.36$\,fb   & $(1.83)0.46$\,\% & $1.12$\,fb & $0.40$\,\%\\\hline
$350$\,GeV & $(1.82)1.71$\,fb   & $(7.77)1.82$\,\% & $1.91$\,fb & $0.88$\,\%\\\hline
$500$\,GeV & $(0.981)0.802$\,fb & $(24.1)7.20$\,\% & $4.78$\,fb & $2.96$\,\%\\\hline
$1$\,TeV   & $(0.341)0.242$\,fb & $(50.9)30.9$\,\% & $15.0$\,fb & $13.0$\,\%\\\hline
  \hline
$\sqrt{s}$ & $\sigma^\zww_{\text{off}}$ & $\Delta_{\text{off}}^\zww$ 
& $\sigma^\nunuww_{\text{off}}$ & $\Delta_{\text{off}}^\nunuww$
\\\hline\hline
$250$\,GeV & $76.3$\,fb & $0.03$\,\% & $(3.99)3.98$\,fb & $(0.12)0.13$\,\%\\\hline
$300$\,GeV & $57.7$\,fb & $0.42$\,\% & $(9.08)9.07$\,fb & $(0.26)0.29$\,\%\\\hline
$350$\,GeV & $41.4$\,fb & $0.92$\,\% & $(15.5)15.5$\,fb & $(0.43)0.49$\,\%\\\hline
$500$\,GeV & $18.6$\,fb & $2.61$\,\% & $(38.1)38.2$\,fb & $(0.96)1.21$\,\%\\\hline
$1$\,TeV   & $4.58$\,fb & $11.0$\,\% & $(108.9)110.8$\,fb & $(2.78)4.45$\,\%\\\hline
\end{tabular}
\end{center}
\vspace{-5mm}
\caption{Inclusive cross sections
$\sigma_{\text{off}}$ for $e^{+}e^{-}\rightarrow ZH\rightarrow ZVV$
and for $e^{+}e^{-}\rightarrow \nu\bar\nu H \rightarrow \nu\bar\nu VV$
for a polarisation of Pol$(e^+,e^-)=(0.3,-0.8)$ and
relative size of the off-shell contributions $\Delta_{\text{off}}$ in \%.
The results averaging over the $ZZ$ pairs for $e^+e^-\rightarrow ZZZ$
and taking into account the $t$-channel Higgs contribution for
$e^+e^-\rightarrow \nu\bar\nu WW$ are added in brackets.}
\label{tab:ZZtable}
\end{table}
As it can be seen from \tab{tab:ZZtable} the off-shell contributions
reach $\mathcal{O}(10\%)$ for large \cms{} energies.
Naturally, the on-shell Higgs cross section is much dependent
on the precise numerical value of the Higgs mass, whereas the
off-shell contributions are insensitive to the latter value.
In extended Higgs sectors the off-shell contributions of the
light \sm{}-like Higgs can interfere with on-shell contributions
of a heavy Higgs, as it was demonstrated for a linear collider in the
context of a 2-Higgs-doublet model in \citere{Liebler:2015aka}.
Similar findings were reported for the \lhc{} in
\citeres{Maina:2015ela}.

The $Z$ recoil mass measurement is the first step of a unique
method to obtain the Higgs width at a linear collider, often named
$Z$ recoil method~\cite{Li:2012taa}.
It is only hardly affected by off-shell
contributions to Higgs boson decays into heavy gauge bosons
at low \cms{} energies $\sqrt{s}=250-350$\,GeV \cite{Liebler:2015aka}.
Contrary off-shell contributions provide their own method of
constraining the Higgs width, when combining them with the
on-shell measurement of Higgs boson decays, however we pointed
to the strong theoretical limitations. To demonstrate
its limitations compared to the $Z$ recoil method
we simulate the process $e^+e^-\rightarrow \nu\bar\nu +4$\,jets
using {\tt MadGraph5\_aMC@NLO}~\cite{Alwall:2011uj} and {\tt MadAnalysis}~\cite{Conte:2012fm}.
For the specific settings of cuts on the external particles
to distinguish the final states and to reduce background
we refer to \citere{Liebler:2015aka}. If at the same time
the Higgs to gauge boson couplings and the total width $\GaH$
entering the on-shell cross section
are varied such, that the on-shell cross section remains
constant, the off-shell contributions develop a dependence
on the Higgs width.
Thus, the total number
of events with an invariant mass of the four jets $\mj>130$\,GeV
can be ultimately expressed in the form~\cite{Liebler:2015aka}
\vspace{-5mm}
\begin{align}
 N(r)=N_0(1+R_1\sqrt{r}+R_2 r)+N_B\quad,
 \label{eq:nrevents}
\end{align}
where $r=\GaH/\GaH^{\sm}$ and $N_0$, $N_B$, $R_1$, $R_2$ are constants.
The linear dependence on $r$ stems from the Higgs boson induced off-shell contributions,
whereas the $\sqrt{r}$ dependence is induced from Higgs to
background interference terms.

\begin{figure}[ht]
\begin{center}
\includegraphics[width=0.35\textwidth]{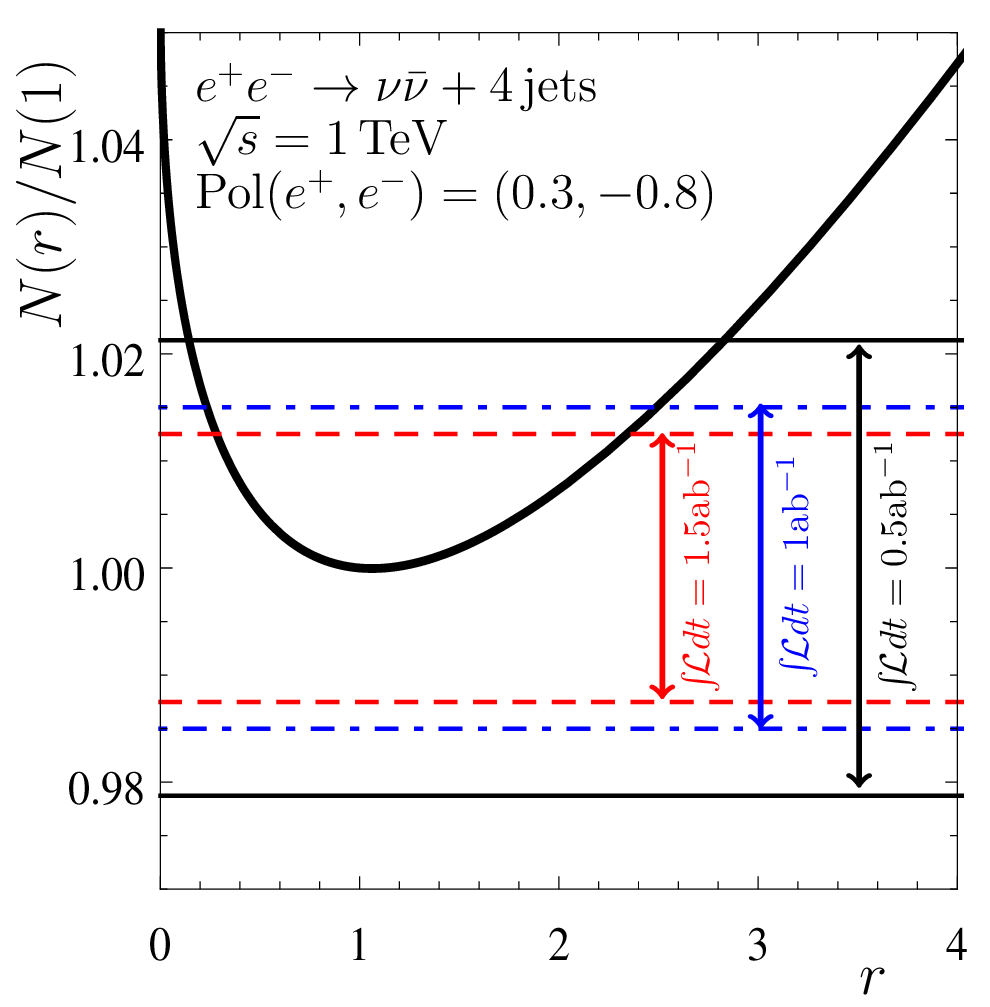} 
\end{center}
\vspace{-0.6cm}
\caption{Normalised event rates $N(r)/N(1)$ as a function of $r$
for $e^+e^-\rightarrow \nu\bar\nu +4$jets for $\sqrt{s}=1$\,TeV
with $95$\% uncertainty bands for different integrated luminosities.}
\label{fig:vvjjjj} 
\end{figure}

We show the normalised event rates $N(r)/N(1)$ for a 
\cms{} energy of $\sqrt{s}=1$\,TeV in \fig{fig:vvjjjj}, where we
add in addition the exclusion range of $r$ and thus $\GaH$
for different integrated luminosities, which
is based on the $95$\% uncertainty band following a simplistic Bayesian
approach~\cite{Liebler:2015aka}.
The sensitivity to $r$ for large $\sqrt{s}$ is even for quite high statistics
lowered by the interference terms of Higgs induced diagrams and background
diagrams, which lead to a minimum of $N(r)$ in the vicinity of one.
A similar reduced sensitivity around $r\sim 1$ applies to the \lhc{} analysis~\cite{Liebler:2015aka}.
At an $e^+e^-$ collider Higgsstrahlung induced processes
are less affected by the negative interference, but
often limited by low statistics.

\section{Signal-background interference in $H\rightarrow \gamma\gamma$}

\begin{figure}[ht]
\begin{center}
\begin{tabular}{cc}
\includegraphics[width=0.3\textwidth]{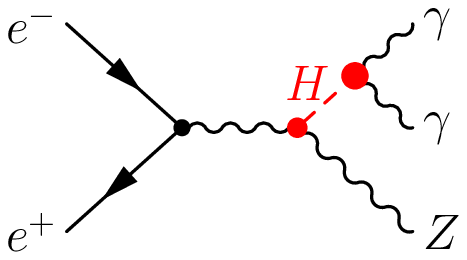} &
\includegraphics[width=0.3\textwidth]{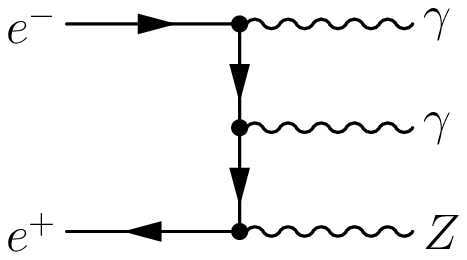} \\[-0.4cm]
 (a) & (b)
\end{tabular}
\end{center}
\vspace{-0.6cm}
\caption{Feynman diagrams for $e^+e^-\rightarrow Z\gamma\gamma$ classified as
(a) signal and (b) background.}
\label{fig:feynmanZgaga} 
\end{figure}

Subsequently we discuss the signal-background interference in $H\rightarrow \gamma\gamma$
at an $e^+e^-$ collider for the Higgsstrahlung process and comment
on the differences for the vector-boson fusion process.
As pointed out in \citere{Martin:2012xc} for the gluon fusion Higgs production
channel at the \lhc{} the mass peak in $H\rightarrow \gamma\gamma$ is shifted
from interference with background, which is meanwhile worked out in more
detail and at higher orders \cite{deFlorian:2013psa}.
A similar shift is also expected at an $e^+e^-$ collider as we demonstrate below.
Again we performed a calculation using {\tt FeynArts} and {\tt FormCalc}~\cite{Hahn:1998yk}
at lowest order and implemented
the Higgs-photon-photon interaction as effective vertex given by~\cite{Martin:2012xc}
\begin{align}
\label{eq:effvertex}
&\parbox{25mm}{\includegraphics[height=0.12\textwidth]{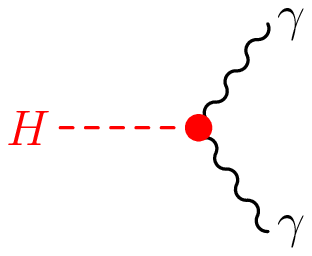}}=
\frac{i\sqrt{\sqrt{2}G_F}}{4\pi}\mgaga^2
\left[F_1(4\mw^2/\mgaga^2)+\sum_{f=t,b,c,\tau} N_f e_f^2 F_{1/2}(4m_f^2/\mgaga^2)\right]\quad,
\end{align}
where $\mgaga^2$ determines the invariant mass of the photons, $N_f=3(1)$
for quarks (leptons) with electric charge $e_f$ and mass $m_f$ and
$G_F$ denotes the Fermi constant.
The functions $F_1$ and $F_{1/2}$ can be found in \citere{Martin:2012xc}.
Making use of the effective vertex in \eqn{eq:effvertex}
example Feynman diagrams for the signal and the background for
$e^+e^-(\rightarrow ZH)\rightarrow Z\gamma\gamma$ are shown in \fig{fig:feynmanZgaga}.

\begin{figure}[ht]
\begin{center}
\begin{tabular}{ccc}
\includegraphics[width=0.3\textwidth]{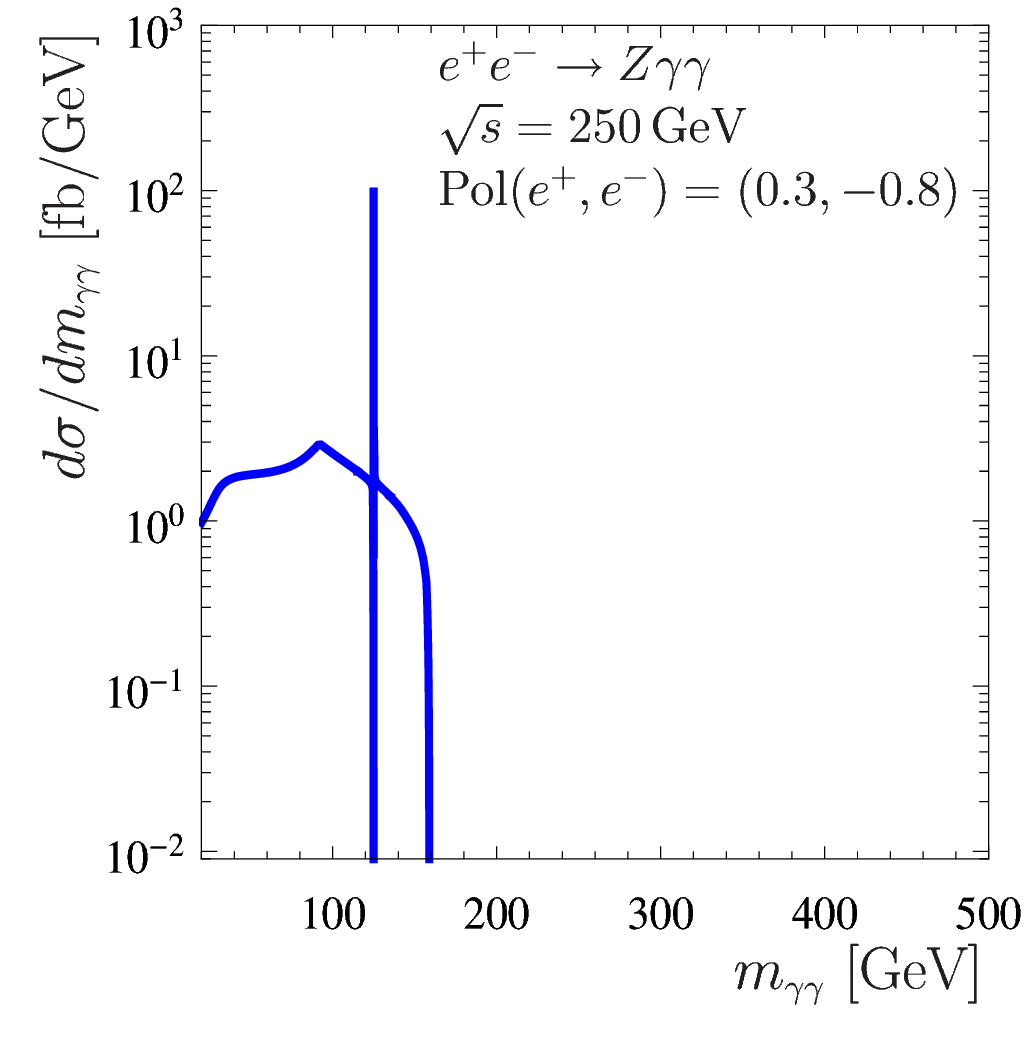} &
\includegraphics[width=0.3\textwidth]{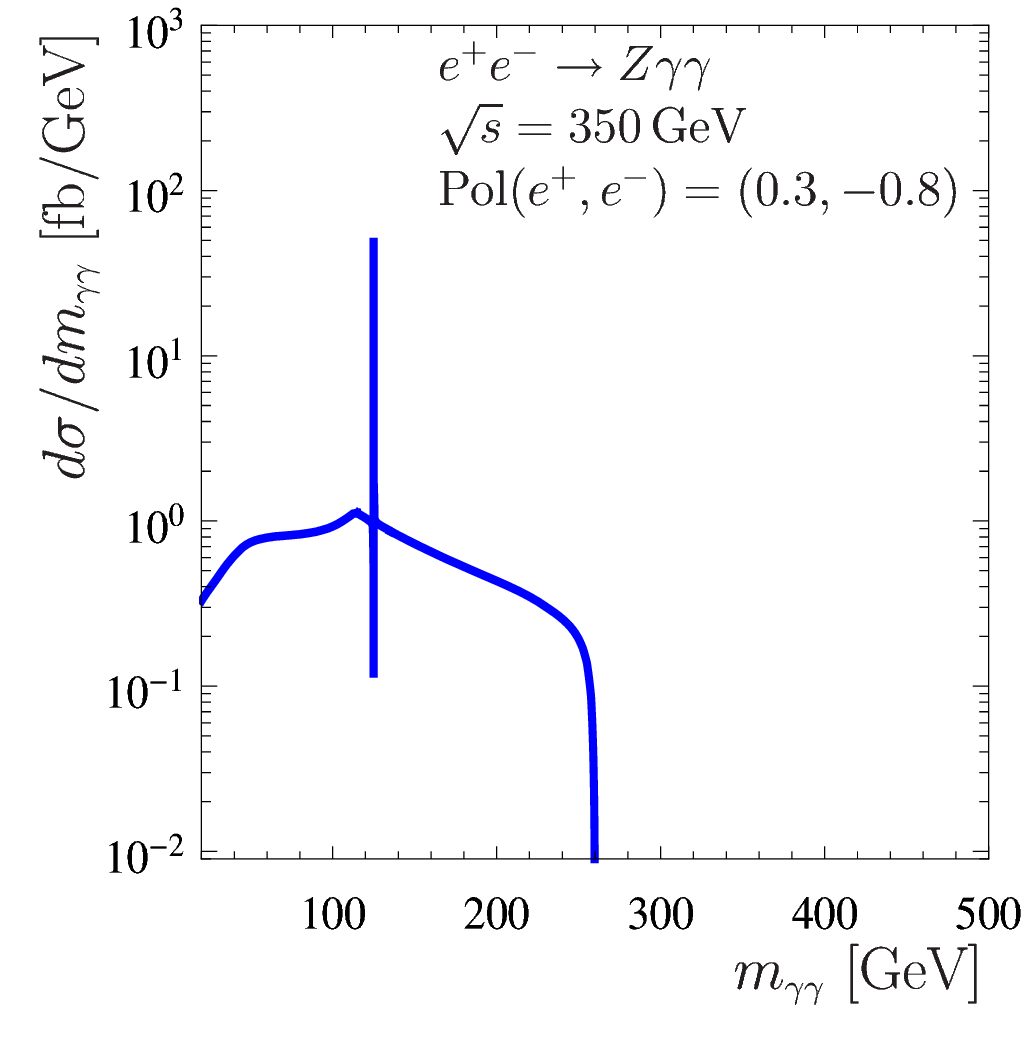} &
\includegraphics[width=0.3\textwidth]{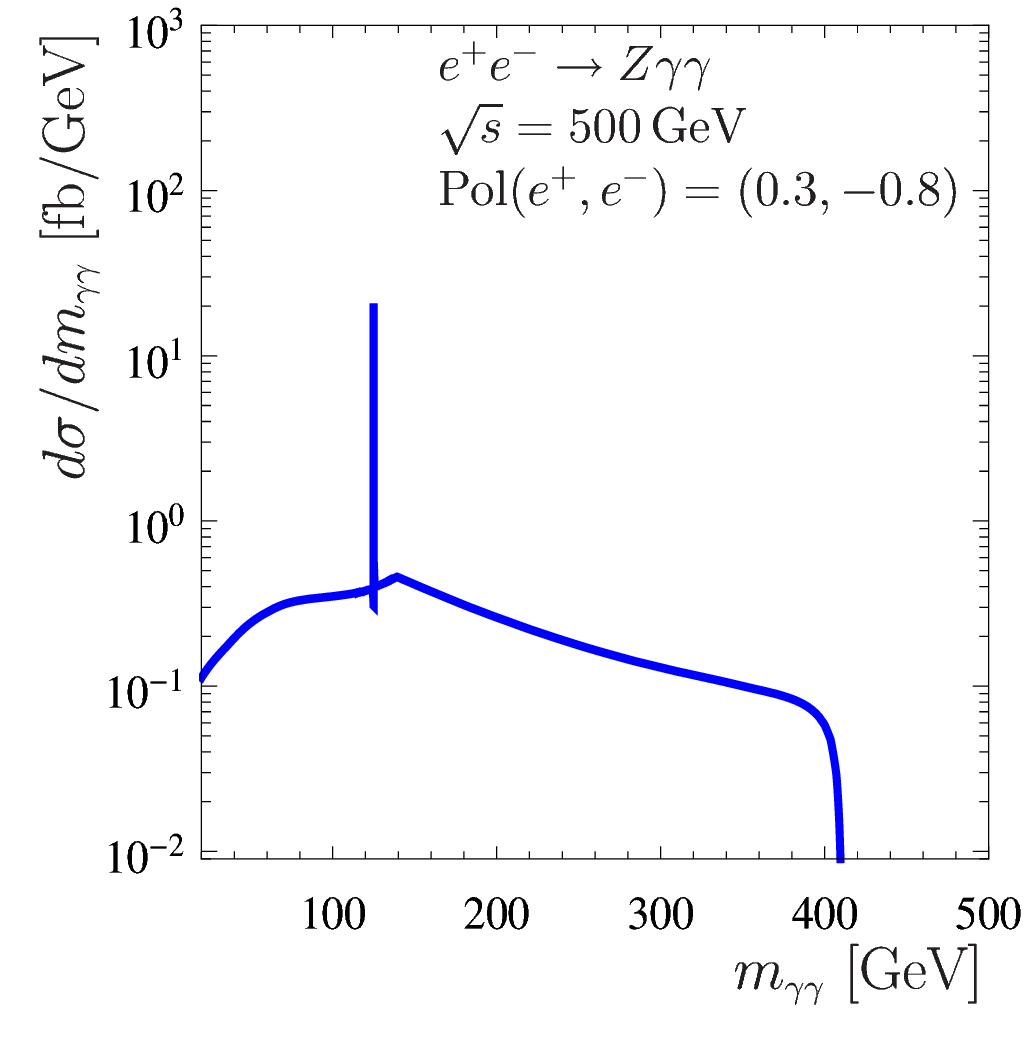} \\[-0.5cm]
 (a) & (b) & (c)
\end{tabular}
\end{center}
\vspace{-0.6cm}
\caption{$d\sigma/d\mgaga$ in fb/GeV as a function of $\mgaga$ in GeV for 
$e^+e^-\rightarrow Z\gamma\gamma$ with \cms{} energies (a-c) $\sqrt{s}=250,\,350,\, 500$\,GeV
for Pol$(e^+,e^-)=(0.3,-0.8)$ and $E_\gamma > 20$\,GeV, $|\eta_\gamma|<2$.}
\label{fig:ZgagaSBI} 
\end{figure}

We start with the discussion of the background for both production processes.
In order to avoid infrared singularities we apply a cut on the photon energies
of $E_\gamma > 20$\,GeV and to take care of collinear singularities we cut
on the photon pseudorapidity $|\eta_\gamma|<2$. Similar cuts were already
applied in previous works~\cite{Barger:1988fd,Stirling:1999ek}.
A cut on the transverse momentum $p_T$ leads to similar findings.
For $e^+e^-\rightarrow Z\gamma\gamma$ \fig{fig:ZgagaSBI} (a-c) show the cross section as a function
of the invariant mass of the two outgoing photons $\mgaga$ in fb/GeV
for \cms{} energies $\sqrt{s}=250,\,350,\,500$\,GeV.
The integrated inclusive cross section is in accordance to \citere{Stirling:1999ek}
and a calculation with {\tt CalcHEP}~\cite{Belyaev:2012qa}.
The differential cross section is reproduced with the help of {\tt MadGraph5\_aMC@NLO}~\cite{Alwall:2011uj}.
In case of $e^+e^-\rightarrow ZH$ the Higgs is
likely to have a non-vanishing transverse momenta, which leads to two photons
with a small opening angle.
Thus, the background and therefore also the signal-background interference is lowered,
when the maximal value of $|\eta_\gamma|$ is increased and/or the opening angle between
the two photons is restricted to small values.

The background around $\mgaga\approx \mh$ follows a rather smooth
behaviour, which allows its subtraction by a side-band analysis and therefore the
determination of the mass peak and its shift due to the signal-background interference.
The maximum in $d\sigma_B^{\zgaga}/d\mgaga$ close to $\mgaga\sim \mh$
stems from the applied cuts and shifts by lowering/increasing the minimal
photon energy $E_\gamma$.
Cross sections for the pure background contribution, named $\sigma_B$, and the signal contribution $\sigma_S$
can be taken from \tab{tab:AAbackground}. The contributions of the interference term $\sigma_I$
to the inclusive cross section are negligible.
All contributions scale equally with the polarisation
of the initial state in case of $e^+e^-(\rightarrow ZH)\rightarrow Z\gamma\gamma$,
thus the relative ratio between signal and background is independent of the polarisation. The same applies
to $e^+e^-(\rightarrow \nu\bar\nu H) \rightarrow \nu\bar\nu \gamma\gamma$ for large $\sqrt{s}$,
where the $Z\rightarrow \nu\bar\nu$ induced final states are of minor relevance.

\begin{table}[htb]
\begin{center}
\begin{tabular}{| c || c | c | c |}
\hline
$\sqrt{s}$ & $\sigma^{\zgaga}_B$ & $\sigma^{\zgaga}_{B,\text{cut}}$ & $\sigma^{\zgaga}_S$ \\\hline
$250$\,GeV & $260$\,fb  & $51.10$\,fb  & $0.89$\,fb\\\hline
$300$\,GeV & $194$\,fb  & $38.88$\,fb   & $0.67$\,fb\\\hline
$350$\,GeV & $152$\,fb  & $29.76$\,fb  & $0.48$\,fb\\\hline
$500$\,GeV & $87$\,fb   & $11.90$\,fb  & $0.21$\,fb\\\hline
$1$\,TeV   & $30$\,fb   & $1.75$\,fb   & $0.05$\,fb\\\hline
\end{tabular}
\end{center}
\vspace{-0.6cm}
\caption{Inclusive cross sections for $e^+e^-(\rightarrow ZH)\rightarrow Z\gamma\gamma$
separated in signal and background contributions for Pol$(e^+,e^-)=(0.3,-0.8)$ and for different 
\cms{} energies $\sqrt{s}$. The background
in the ``signal region'' is defined by
$\sigma_{B,\text{cut}}=\sigma_{B}(110\,\text{GeV},140\,\text{GeV})$
in the notation of \eqn{eq:uplowsigma}.}
\label{tab:AAbackground}
\end{table}

\begin{figure}[ht]
\begin{center}
\begin{tabular}{ccc}
\includegraphics[width=0.3\textwidth]{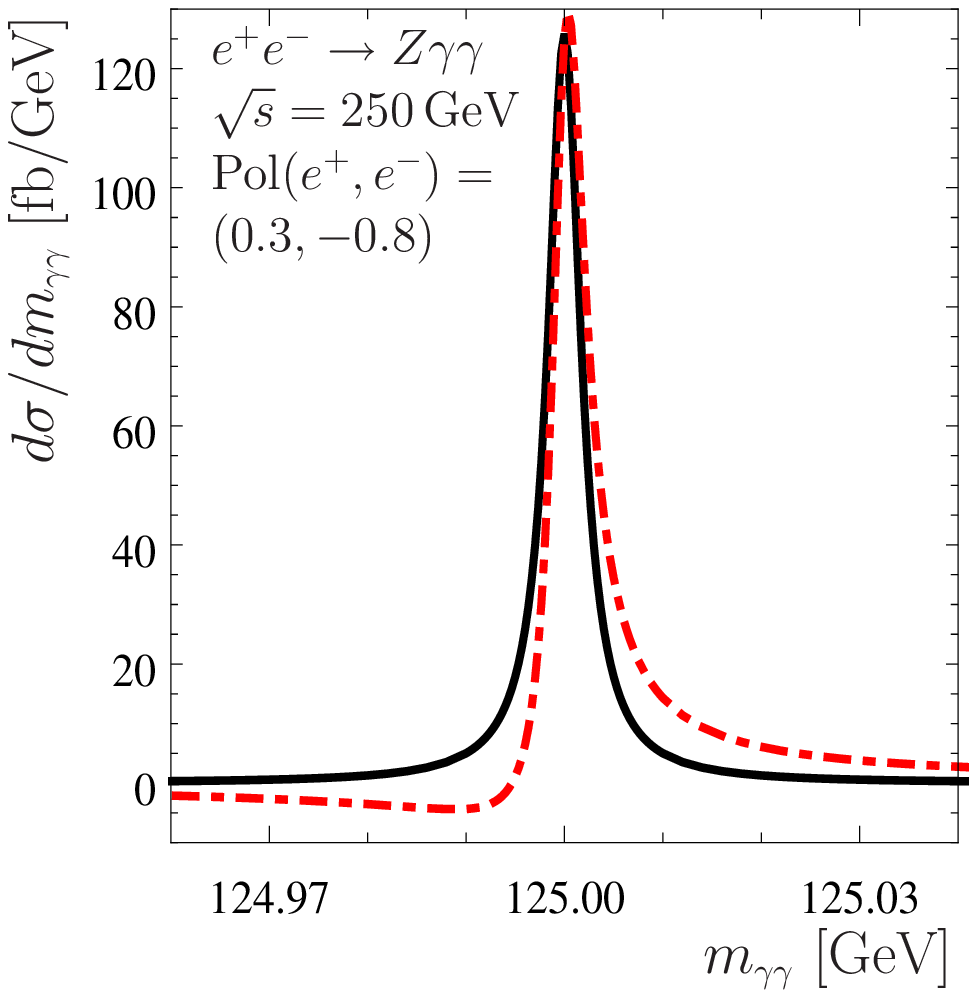} &
\includegraphics[width=0.3\textwidth]{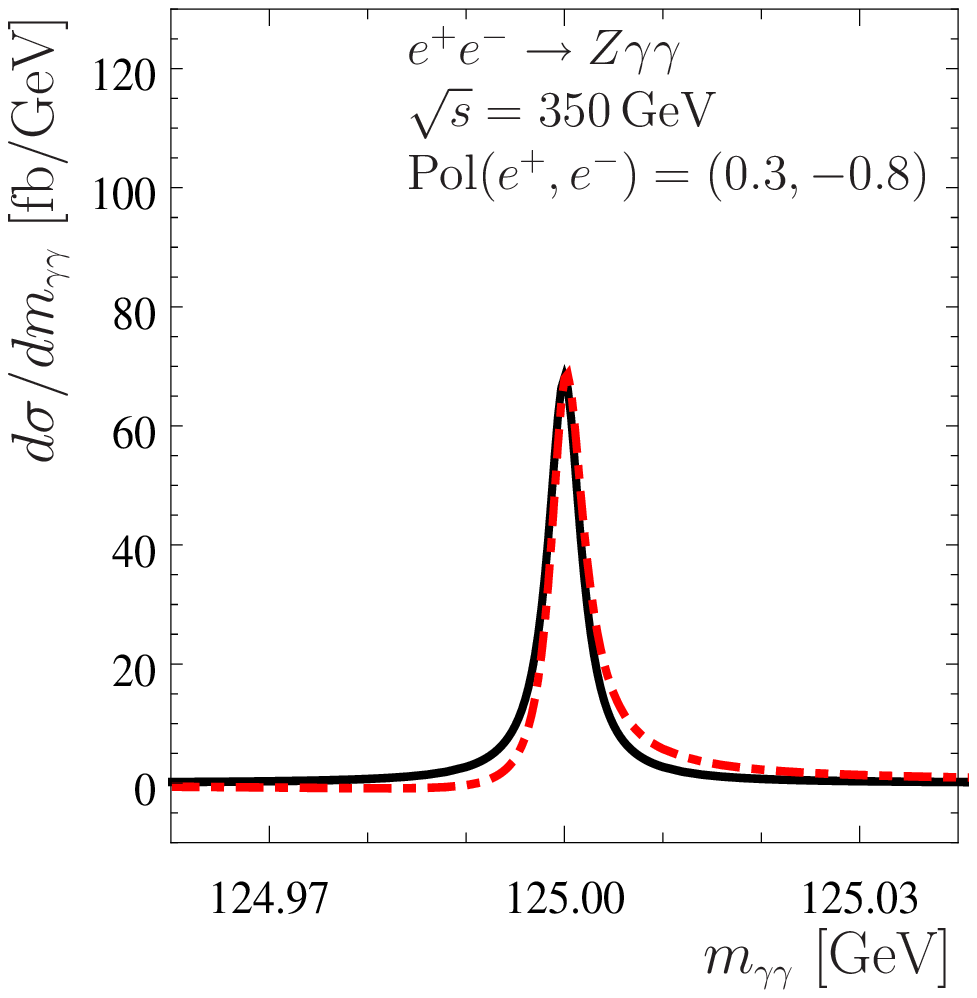} &
\includegraphics[width=0.3\textwidth]{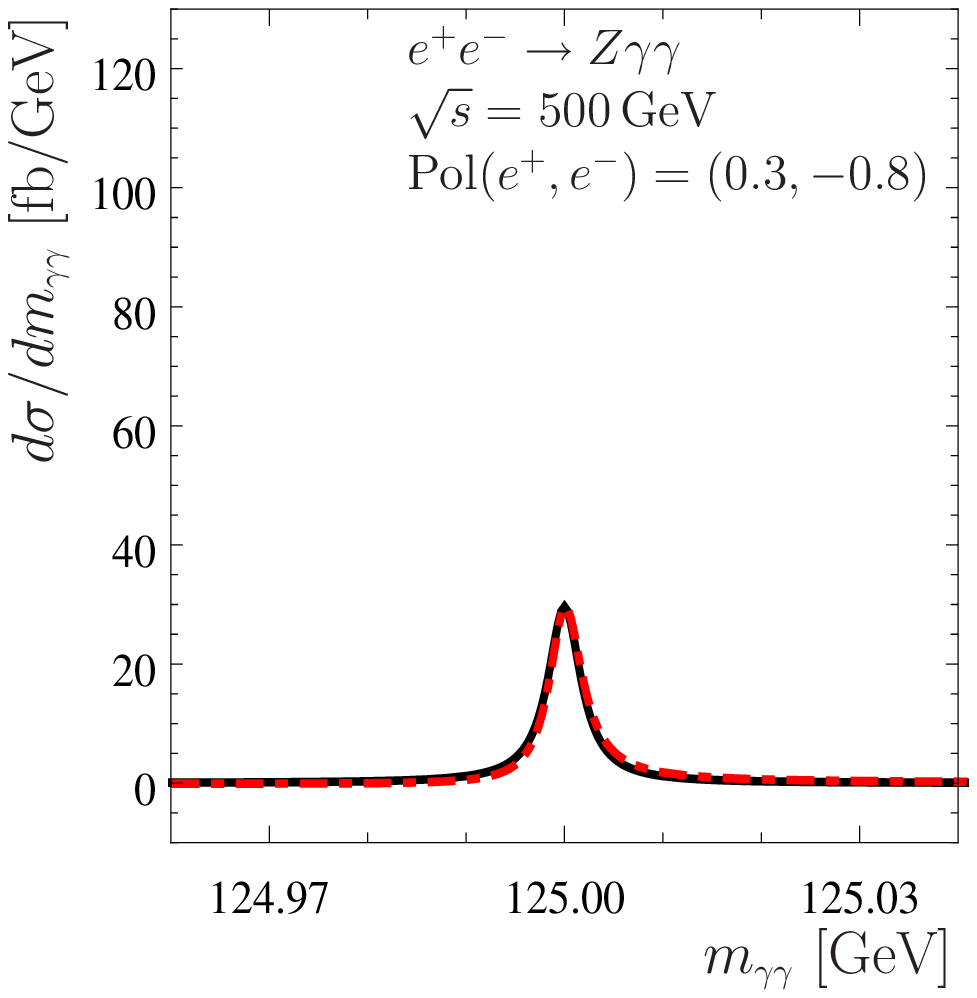} \\[-0.5cm]
 (a) & (b) & (c)
\end{tabular}
\end{center}
\vspace{-0.6cm}
\caption{$d\sigma_S/d\mgaga$ (black, solid) and $\sigma_{S+I}/d\mgaga$ (red, dot-dashed)
in fb/GeV as a function of $\mgaga$ in GeV for
$e^+e^-\rightarrow Z\gamma\gamma$ for (a-c) $\sqrt{s}=250,\,350,\, 500$\,GeV.}
\label{fig:ZgagaSI} 
\end{figure}

\begin{figure}[ht]
\begin{center}
\begin{tabular}{ccc}
\includegraphics[width=0.3\textwidth]{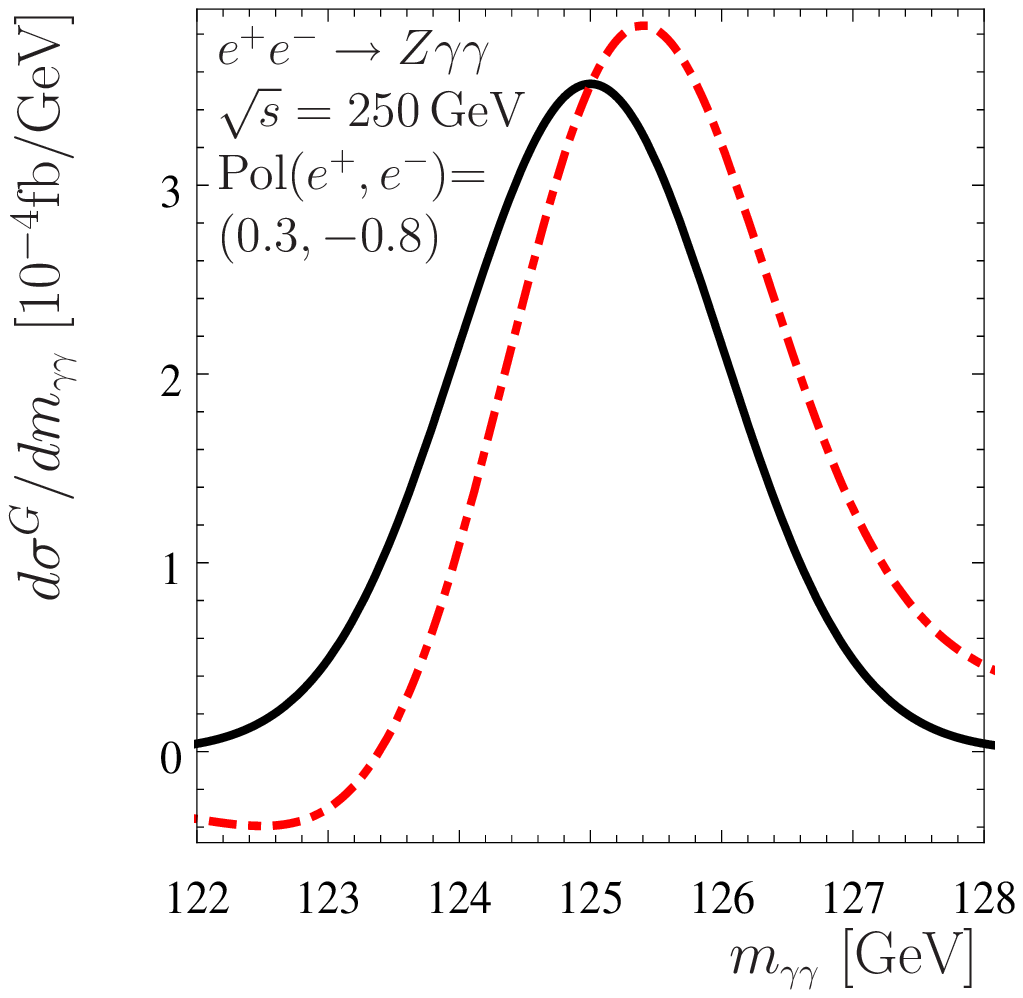} &
\includegraphics[width=0.3\textwidth]{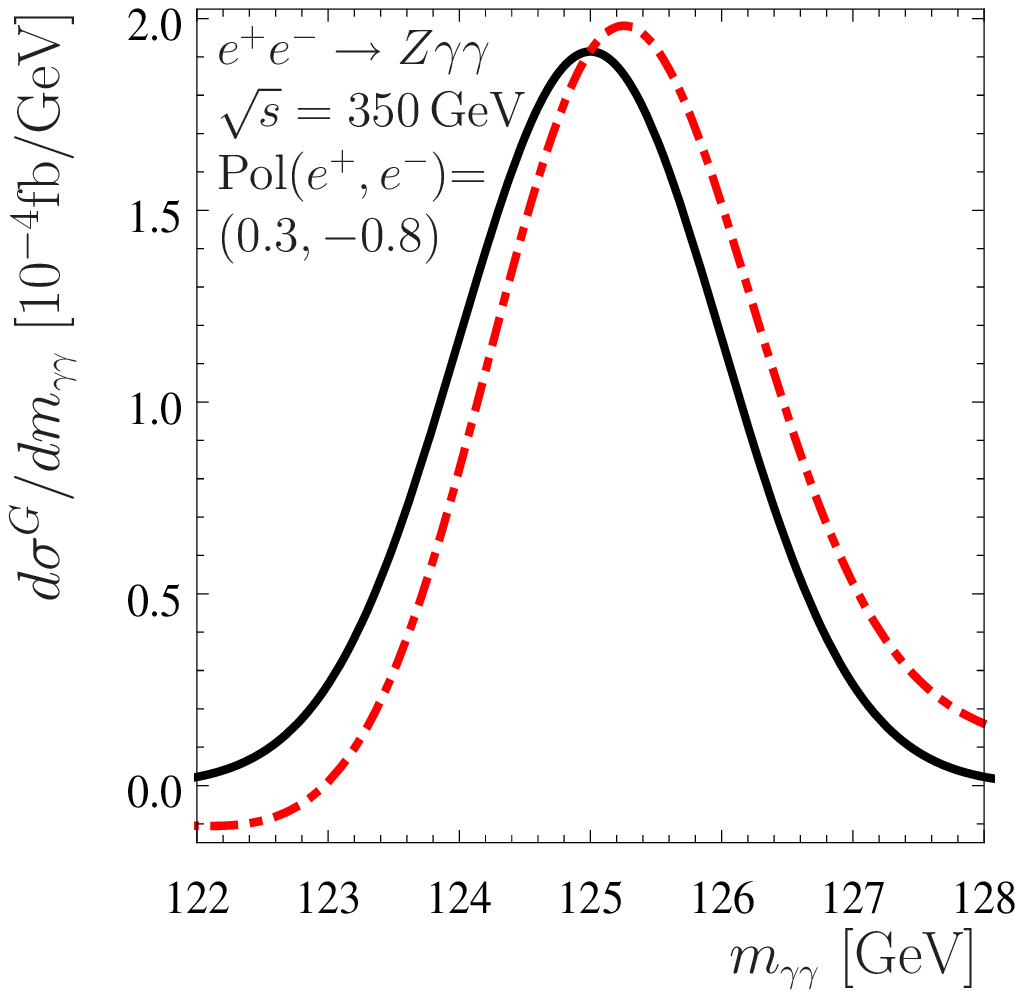} &
\includegraphics[width=0.3\textwidth]{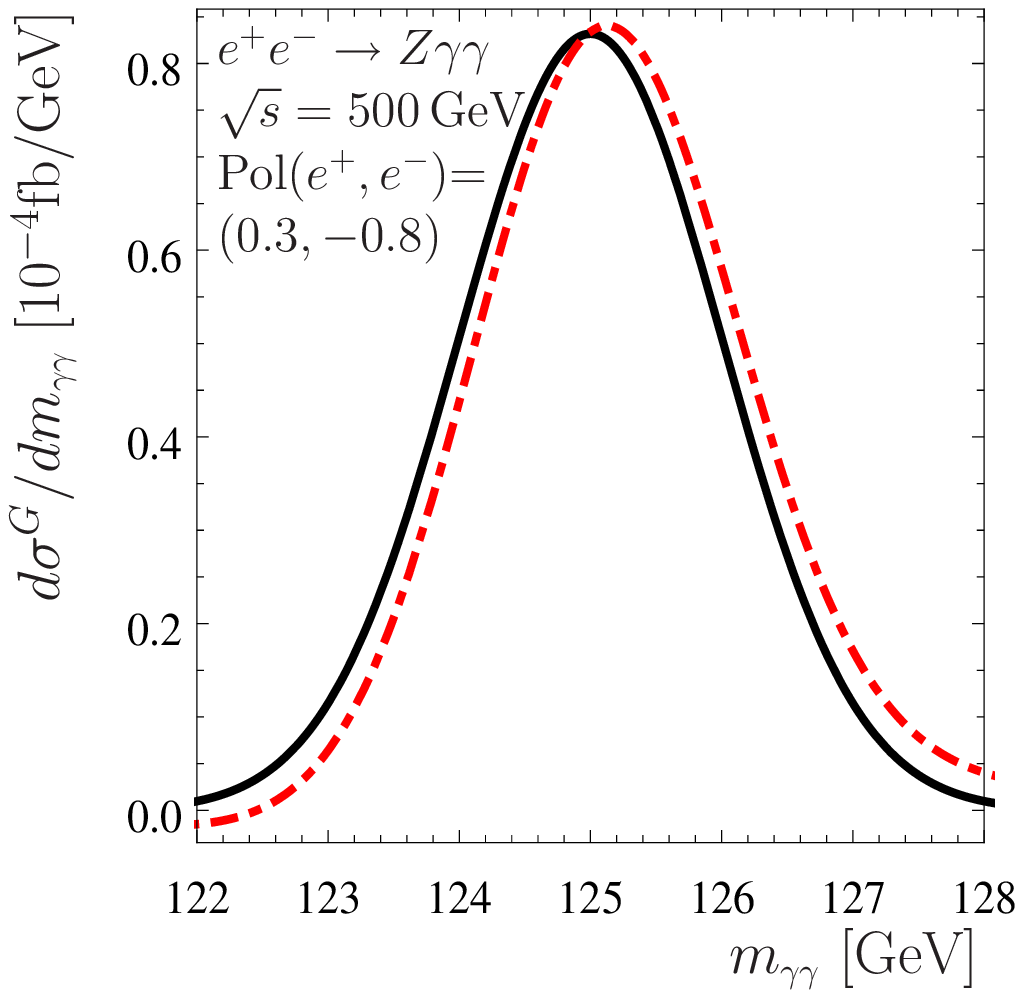} \\[-0.5cm]
 (a) & (b) & (c)
\end{tabular}
\end{center}
\vspace{-0.6cm}
\caption{Smeared $d\sigma^G_S/d\mgaga$ (black, solid) and $\sigma^G_{S+I}/d\mgaga$ (red, dot-dashed)
 in fb/GeV as a function of $\mgaga$ in GeV for
$e^+e^-\rightarrow Z\gamma\gamma$ with $\hat{\sigma}=1$\,GeV for (a-c) $\sqrt{s}=250,\,350,\, 500$\,GeV.}
\label{fig:ZgagaSIsmeared} 
\end{figure}

In \fig{fig:ZgagaSI} we focus on the signal contribution $\sigma_S$, being
the process $e^+e^-\rightarrow ZH\rightarrow Z\gamma\gamma$, and 
the signal-background interference $\sigma_{S+I}$ in the window
around $\mgaga\approx 125$\,GeV using the described cuts and polarisation
of the initial state. The photon energies are usually smeared by
detector effects, which thus need to be discussed by the experimental collaborations.
We convoluted with a Gaussian function with a Gaussian width
of $\hat{\sigma}=1$\,GeV.
The resulting cross sections are denoted $\sigma^G$. 
This convolution lowers the height of $\sigma_S\rightarrow \sigma^G_S$
and $\sigma_{S+I}\rightarrow \sigma^G_{S+I}$, but broadens the peaks accordingly. The
mass shift is therefore enlarged. For the case of \fig{fig:ZgagaSI}
the smeared mass peaks are shown in \fig{fig:ZgagaSIsmeared}. In case
$Z$ boson decays are considered in addition, more background diagrams
get involved and thus affect the shift of the mass peak.


\begin{figure}[ht]
\begin{center}
\begin{tabular}{cc}
\includegraphics[width=0.3\textwidth]{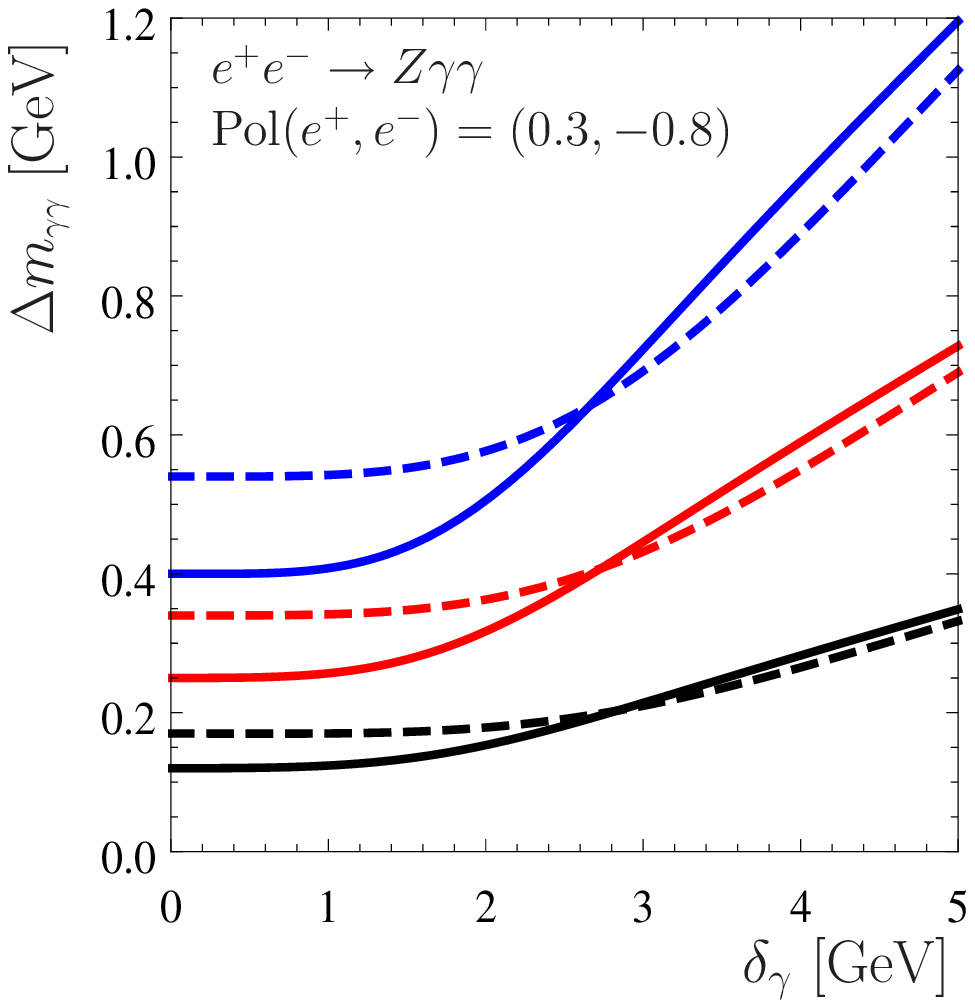} &
\includegraphics[width=0.3\textwidth]{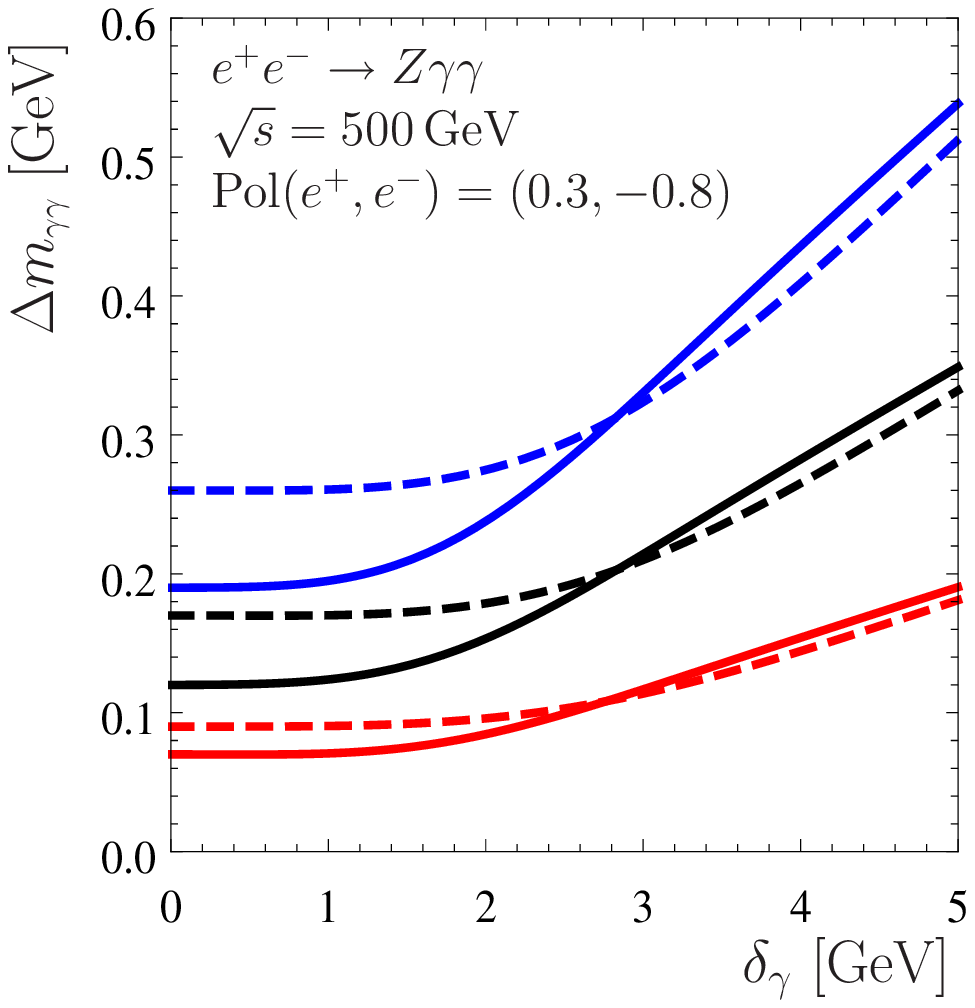}\\[-0.5cm]
 (a) & (b)
\end{tabular}
\end{center}
\vspace{-0.6cm}
\caption{(a) $\Delta \mgaga$ in GeV as a function of $\delta_\smallga$ in GeV 
for $e^+e^-\rightarrow Z\gamma\gamma$ with \cms{} energies $\sqrt{s}=250$\,GeV (blue),
$350$\,GeV (red) and $500$\,GeV (black);
(b) $\Delta \mgaga$ in GeV as a function of $\delta_\smallga$ in GeV 
for $e^+e^-\rightarrow Z\gamma\gamma$ for $\sqrt{s}=500$\,GeV with $\GaH=1$\,MeV (red),
$4.07$\,MeV (black) and $15$\,MeV (blue).
Both figures include two Gaussian widths $\hat{\sigma}=1$\,GeV (solid) and $\hat{\sigma}=1.5$\,GeV (dashed).
}
\label{fig:Deltamgaga} 
\end{figure}

To quantify the shift we closely follow \citere{Martin:2012xc}. Supposing that
the experimental collaborations reduce the background by a suitable side-band
analysis, we are left with the shifted signal event rate. To determine
the mass peak a simplistic way is given by the mean $\langle\mgaga\rangle$
within the interval $[m_p-\delta_\smallga,m_p+\delta_\smallga]$, where $m_p$ is the actual
mass peak of the distribution and $\delta_\smallga$ is given by the experiment.
Then we define \cite{Martin:2012xc}
\begin{align}
\langle \mgaga\rangle_{\delta,X} = \frac{1}{N_{\delta,X}}
\int_{m_p-\delta}^{m_p+\delta} d\mgaga \mgaga \frac{d\sigma^G_X}{d\mgaga}
\qquad\text{with}\qquad
N_{\delta,X} = \int_{m_p-\delta}^{m_p+\delta} d\mgaga \frac{d\sigma^G_X}{d\mgaga}\quad.
\end{align}
A theoretical measure of the shift is given by
\begin{align}
\Delta \mgaga = \langle \mgaga\rangle_{\delta_\smallga,S+I} - \langle \mgaga\rangle_{\delta_\smallga,S} \quad.
\end{align}
We note that the experiments can access only $\langle \mgaga\rangle_{\delta,S+I}$.
To obtain a normalization of the shift, either other production or decay channels
have to be considered to obtain the actual Higgs mass or alternatively different
kinematic regions in $H\rightarrow\gamma\gamma$ allow for the observation of different shifts.
The latter method can e.g. be understood by applying different cuts on the transverse momentum
$p_T$ or the angle between the two photons. A detailed detector simulation is on order, which
is beyond the scope of our considerations.
We instead try to quantify the effect using the theoretical measure $\Delta \mgaga$
as a function of $\delta_\smallga$ and present $\Delta \mgaga$ for two Gaussian
widths $\hat{\sigma}=1$ and $1.5$\,GeV in \fig{fig:Deltamgaga}~(a)
for different \cms{} energies.

The large background for $e^+e^-\rightarrow Z\gamma\gamma$ at low \cms{} energies
$\sqrt{s}$ induces a sizeable shift~$\Delta \mgaga$.
For low values of $\delta_\smallga$ the mass shift $\Delta \mgaga$ corresponds to the
actual difference between the two peaks of the distributions after Gaussian smearing
as given in \fig{fig:ZgagaSIsmeared}.
For large values of $\delta_\smallga>2\hat{\sigma}$ the mass shift $\Delta \mgaga$ increases
linearly, since more of the tail of the signal-background interference is included.
To show the dependence of the shift on the width $\GaH$ of the Higgs boson, we perform
a similar analysis as in \sct{sec:offshell} and change the Higgs width in combination with
a rescaling of the $HVV$ couplings to leave the inclusive on-shell Higgs cross section constant.
For simplicity we rescale all $HVV$ couplings including the loop-induced Higgs boson to photons
coupling equally.
Apart from the \sm{} value we pick the Higgs widths $\GaH=1$\,MeV and $\GaH=15$\,MeV.
\fig{fig:Deltamgaga}~(b) shows $\Delta\mgaga$ as a function of $\delta_\smallga$ for
different values of $\GaH$ for $\sqrt{s}=500$\,GeV.
The mass shift, being of the order of $\mathcal{O}(100$\,MeV), shows a clear dependence
on the Higgs width.
Similar findings and shifts in the invariant mass peak of the two photons
can be obtained for the vector-boson fusion process, but we leave a detailed presentation
to future work.

\section{Conclusions}
\label{sec:conclusions}

We quantified the shift of the invariant mass peak
in Higgs boson to photon decays $H\rightarrow \gamma\gamma$
induced by the signal-background interference at an $e^+e^-$ collider for
the Higgsstrahlung process $e^+e^-\rightarrow Z\gamma\gamma$. The observed mass shift
is strongly dependent on detector effects, but expected to be in the range of
$\mathcal{O}(100$\,MeV). Similar effects occur for the vector-boson fusion induced
Higgs production process and both allow to access the Higgs width.
Similarly we quantified off-shell contributions in $H\rightarrow VV^{(*)}$ with $V\in\lbrace Z,W\rbrace$
for both production processes for different polarisations and \cms{} energies.
At larger \cms{} energies $\sqrt{s}>500$\,GeV they contribute $\mathcal{O}(10\%)$ to the total Higgs
boson induced cross section. Whereas the $Z$ recoil method at low \cms{} energies
is thus safe from off-shell contributions, the off-shell region allows to
extract the Higgs boson to gauge boson couplings in different kinematical regimes
and thus to test the electroweak symmetry breaking mechanism.
The determination of the Higgs width from a combination of on- and off-shell
measurements in Higgs boson decays to heavy gauge bosons
-- besides relying on strong theoretical assumptions -- is mainly
limited by the negative signal-background interference term in the off-shell region.

\section*{Acknowledgments}

The author acknowledges support by
``Deutsche Forschungsgemeinschaft'' through the SFB~676 ``Particles, Strings and the Early Universe''.


\end{document}